\documentclass[superscriptaddress,twocolumn,
nobibnotes,aps,prd,showpacs,nofootinbib]{revtex4}
\usepackage{graphicx}
\usepackage{epsf}
\usepackage{bm}
\usepackage{amsmath}
\usepackage{amsfonts}
\usepackage{amssymb}
\usepackage{epstopdf}
\usepackage{natbib}
\usepackage{color}
\usepackage{color}
\providecommand{\U}[1]{\protect\rule{.1in}{.1in}}
\newcommand{\be}{\begin{equation}}
\newcommand{\ee}{\end{equation}}

\newcommand{\mincir}{\raise
-3.truept\hbox{\rlap{\hbox{$\sim$}}\raise4.truept\hbox{$<$}\ }}
\newcommand{\magcir}{\raise
-3.truept\hbox{\rlap{\hbox{$\sim$}}\raise4.truept\hbox{$>$}\ }}

\begin{document}

\title{Observational constraints on dynamical dark energy with pivoting redshift}

\author{Weiqiang Yang}
\email{d11102004@163.com}
\affiliation{Department of Physics, Liaoning Normal University, Dalian, 116029, P. R.
China}

\author{Supriya Pan}
\email{supriya.maths@presiuniv.ac.in}
\affiliation{Department of Mathematics, Presidency University, 86/1 College Street, Kolkata 700073, India}

\author{Eleonora Di Valentino}
\email{eleonora.divalentino@manchester.ac.uk}
\affiliation{Jodrell Bank Center for Astrophysics, School of Physics and Astronomy, University of Manchester, Oxford Road, Manchester, M13 9PL, UK}

\author{Emmanuel N. Saridakis}
\email{Emmanuel_Saridakis@baylor.edu}

\affiliation{Department of Physics, National Technical University of Athens, Zografou
Campus GR 157 73, Athens, Greece}
\affiliation{CASPER, Physics Department, Baylor University, Waco, TX 76798-7310,
USA}

\begin{abstract}
We investigate the generalized Chevallier-Polarski-Linder (CPL) parametrization, 
which contains the pivoting redshift $z_p$ as an extra free parameter. We 
use various data combinations  
from cosmic microwave background (CMB), 
baryon acoustic oscillations (BAO),
redshift space distortion (RSD), 
 weak lensing (WL), joint light curve analysis (JLA), 
cosmic chronometers (CC), and we include a Gaussian prior on the Hubble constant value,
in order to extract the observational constraints on 
various quantities.  
For the case of free $z_p$ we find that for all data combinations it always remains unconstrained, and there is a degeneracy with the current 
value of the dark energy equation of state $w_0$. For the case where $z_p$ is fixed to specific values, and for the full data combination,
we find that with increasing $z_p$ the mean value of $w_0$  
slowly moves into the phantom regime, however the  cosmological constant is always allowed within 1$\sigma$ confidence-level.
However, the significant effect is that
with increasing $z_p$ the correlations between $w_0$ and $w_a$
change from negative to positive, with the case $z_p =0.35$ corresponding to no correlation.
This feature indeed justifies why a non-zero pivoting redshift should be taken into account.
\end{abstract}

\maketitle


\section{Introduction} 

According to observations the universe has entered a period of accelerated expansion in the recent cosmological past. 
In order to provide an explanation, physicists follow two main directions. The first is to maintain general relativity as the gravitational theory and introduce new, exotic fluids in the universe content, dubbed as dark energy sector  \cite{Copeland:2006wr,Cai:2009zp}. The second way is to modify the gravitational sector, constructing extended theories of gravity that possess general relativity as a particular limit but which in general present extra degrees of freedom,  capable of describing the universe behavior \cite{modgrav1, DeFelice:2010aj,Capozziello:2011et,Cai:2015emx,Nojiri:2017ncd}.

However, both approaches can be quantified by the introduction of an equation-of-state 
parameter for the dark energy perfect fluid (effective in the case of modified gravity), namely $w_x(z) = p_x(z)/\rho_x(z)$, where $p_x(z)$, $\rho_x(z)$ are respectively the pressure and the energy density. Hence, introducing various parametrizations of $w_x(z)$ allows us to describe the universe evolution in a phenomenological way, even if the microphysical origin of the cause of acceleration is unknown.
Following this, a large number of parametrizations have 
been introduced in the last years, namely
the one-parameter dark energy parametrizations \cite{Gong:2005de,Yang:2018qmz}, or
the two-parameters family, such as the
Chevallier-Polarski-Linder (CPL) parametrization 
\cite{Chevallier:2000qy,Linder:2002et}, 
the Linear parametrization 
\cite{Cooray:1999da,Astier:2000as,Weller:2001gf}, the
Logarithmic parametrization \cite{Efstathiou:1999tm}, the Jassal-Bagla-Padmanabhan 
parametrization (JBP) \cite{Jassal:2005qc},  the
Barboza-Alcaniz (BA) parametrization \cite{Barboza:2008rh}, 
etc (see for instance \cite{Ma:2011nc,Nesseris:2004wj, Linder:2005ne, Feng:2004ff, Zhao:2006qg, Nojiri:2006ww, Saridakis:2008fy,Dutta:2009yb,Lazkoz:2010gz, Feng:2011zzo,Saridakis:2009pj, DeFelice:2012vd,Saridakis:2009ej, Feng:2012gf,Basilakos:2013vya, Pantazis:2016nky, DiValentino:2016hlg, Chavez:2016epc,Zhao:2017cud, Yang:2017amu, DiValentino:2017zyq, DiValentino:2017gzb, Yang:2017alx, Marcondes:2017vjw, Pan:2017zoh} and  references therein). 

In most of the above dark-energy equation-of-state parametrizations one considers the ``pivoting redshift'' to
correspond to zero, namely the  point in which $w_{x}$ is most tightly constrained to correspond to the current universe. However, due to possible rotational correlations between the two parameters of the
two-parameter models, in principle one could avoid setting the pivoting redshift to zero
straightaway, and let it as a free parameter  \cite{Linder:2006xb,Sahni:2006pa}.  

In the present work we are interested in investigating the observational constraints on the 
most well-known parametrization, namely the CPL one, incorporating however the 
pivoting redshift as an extra parameter, assuming it to be either fixed or free.
A first examination towards this direction was
performed in  \cite{Scovacricchi:2014haa}, however in the present work we provide a robust 
analysis with the latest cosmological data. In particular, we will use data 
from cosmic microwave background (CMB), 
baryon acoustic oscillations (BAO),
redshift space distortion (RSD), weak lensing (WL), joint light curve analysis (JLA), 
cosmic chronometers (CC), while we will include a Gaussian prior on the Hubble constant value.
 
The plan of the work is the following: In Section \ref{sec:2} we  present the basic equations
of a parametrized dark energy model at the background and perturbative levels. Section
\ref{sec-data} describes the various data sets used in this work.
In Section \ref{sec-results} we perform the observational confrontation, 
extracting the constraints on the model parameters and on various cosmological quantities. 
Finally, we summarize the obtained results   in Section \ref{sec-conclu}.

\section{Dynamical Dark-Energy with pivoting redshift}
\label{sec:2}

In this section we briefly review the basic  equations for a non-interacting cosmological scenario both at background and perturbative levels, and we   introduce the pivoting redshift dark energy parametrization. We consider 
the homogeneous and isotropic Friedmann-Lema\^itre-Robertson-Walker (FLRW) line element 
\begin{eqnarray}
 ds^2 = - dt^2 + a^2(t) \left[ \frac{ dr^2}{1-Kr^2} + r^2 \left ( 
d\theta^2 
+ \sin^2 
\theta d \phi^2   \right)  \right]\!,
\end{eqnarray} 
where $a(t)$ is the scale factor and $K = -1,+1,0$  corresponds to
open, closed and flat geometry, respectively. Additionally, we consider that the universe is filled with baryons, cold dark matter, radiation, and the (effective) dark energy fluid. Hence, the evolution of the universe is determined by
the Friedmann equations, which are written as
\begin{eqnarray}
H^2 + \frac{K}{a^2} &=& \frac{8\pi G}{3} \rho_{tot},\label{f1}\\
2\dot{H} + 3 H^2  + \frac{K}{a^2} &=& - 8 \pi G\, p_{tot}\label{f2},
\end{eqnarray}
where  $G$ is the Newton's gravitational constant and $H=\dot{a}/a$ is the Hubble function, with dots 
denoting derivatives with respect to the cosmic time $t$. Moreover, in the above equations we have 
introduced the total energy density and pressure of the universe, reading as
$\rho_{tot} = \rho_r +\rho_b +\rho_c 
+\rho_x$ and $p_{tot} = p_r + 
p_b + p_c + p_x$, with the symbols $r,\; b,\; c,\; x$ corresponding to
radiation, baryon, cold dark matter and   dark energy fluid, respectively.
In the following we focus our analysis on the spatially flat case  ($K=0$), which is the one favored by observations.

In the case where the above sectors  do not present mutual interactions, 
we can write the conservation equation of each fluid  as
\begin{eqnarray}
\label{cons}
\dot{\rho}_i + 3 H (1 +w_i ) \rho_i = 0,
\end{eqnarray} 
where $w_i\equiv p_i/\rho_i$ is known as the equation-of-state parameter
 of the $i$-th fluid ($i \in \{r,\; b,\; c,\; x\}$). In the case of the dark energy fluid, the solution of (\ref{cons})
is
\begin{eqnarray}\label{de-evol}
\rho_{x}=\rho_{x,0}\,\left(  \frac{a}{a_{0}}\right)  ^{-3}\,\exp\left[
-3\int_{a_{0}}^{a}\frac{w_{x}\left(  a'\right)  }{a'}\,da'
\right],
\end{eqnarray}
where $\rho_{x,0}$ is the current value of $\rho_x$ and $a_0$ is the present value of the scale factor
which is set to unity. 

From expression (\ref{de-evol}) we deduce that the evolution of 
the dark energy component is highly dependent on the form of its 
equation-of-state parameter $w_x (a)$. In the simplest case where $w_x (a)=w_0=const.$
 the dark energy fluid evolves as $\rho_x = \rho_{x,0} a^{-3(1+w_0)}$. Nevertheless, 
for dynamical $w_x (a)$ one may consider   various    parametrizations in terms of the scale factor 
or the redshift $z$, where $1+z = a_0/a= 1/a$.
Thus, in the literature one can find many forms of  such parametrizations.

One of the  well known parametrizations of the dark-energy 
equation-of-state parameter is the Chevallier-Polarski-Linder (CPL) one, given by \cite{Chevallier:2000qy,Linder:2002et}
\begin{eqnarray}\label{cpl}
w_x (z) = w_0 + w_a \left(1- a\right),
\end{eqnarray}
where $w_0$ is the current value of $w_x$ and $w_a \equiv dw_{x}/da$ at $a = a_0=1$. 
One can see that introducing an extra parameter,
expression (\ref{cpl}) can  be rewritten as 
\begin{eqnarray}\label{cpl-pivot}
w_{x} (z) = w_0^{p} + w_a^{p} \left( a_p - a \right),
\end{eqnarray}
where $w_0^{p} = w_0 - w_a \left(1-a_p \right)$, $w_a^{p} = w_a$ and $1+z_p = 1/a_p$.
In the case where the extra parameter 
$z_p= 0$, we obtain  $a_p =1$, and thus we recover the standard  CPL model (\ref{cpl}). 
The parameter $z_p$ is called the ``pivoting redshift'' with $a_p$  its corresponding scale factor,
since it marks the point in which $w_{x}$ is most tightly constrained \cite{Linder:2006xb,Sahni:2006pa}.
In particular, it is known that in the above parametrization  $z_p$, and thus  $w_0^{p}$, depend on the 
probing method, the fiducial scenario, and the imposed priors \cite{Linder:2006xb}. Hence, in principle
one could avoid setting $z_p=0$ straightaway, and let it as a free parameter. Thus, $w_0^{p}$ can
be more precisely determined than $w_0$, and actually it is indeed the  
most precisely determined   value of $w_{x} (z)$.  In this work we are interested in investigating the
generalized CPL parametrization (\ref{cpl-pivot}), namely incorporating 
the  pivoting redshift as an extra parameter,
assuming it to be either fixed or free.

We proceed by providing the cosmological equations at the perturbation level.
In the synchronous gauge  the perturbed FLRW metric  reads as  
\begin{eqnarray}
\label{perturbed-metric}
ds^2 = a^2(\tau) \left [-d\tau^2 + (\delta_{ij}+h_{ij}) dx^idx^j  \right], 
\end{eqnarray}
where $\delta_{ij}$ is the unperturbed and $h_{ij}$  the  perturbed metric,  and $\tau$ 
is the conformal time. Using the above perturbed   metric one can solve the 
conservation equations $T^{\mu \nu}_{; \nu}= 0$.
Thus, for a mode with
wavenumber ${k}$ the perturbed equations can be written as 
\cite{Mukhanov, Ma:1995ey, Malik:2008im}
\begin{eqnarray}
\delta'_{i}  = - (1+ w_{i})\, \left(\theta_{i}+ \frac{h'}{2}\right) -
3\mathcal{H}\left(\frac{\delta p_i}{\delta \rho_i} - w_{i} \right)\delta_i
\nonumber\\- 9 \mathcal{H}^2\left(\frac{\delta p_i}{\delta \rho_i} 
- c^2_{a,i} \right) (1+w_i) \frac{\theta_i}{{k}^2}, \label{per1} \\
\theta'_{i}  = - \mathcal{H} \left(1- 3 \frac{\delta p_i}{\delta \rho_i}
\right)\theta_{i} + \frac{\delta p_i/\delta \rho_i}{1+w_{i}}\, {k}^2\, 
\delta_{i} -{k}^2\sigma_i,
\label{per2}
\end{eqnarray}
where primes denote derivatives with respect to the conformal time, and  $\mathcal{H}= a^{\prime}/a$
is the conformal Hubble factor. Additionally, $\delta_i = \delta \rho_i/\rho_i$ is the 
density perturbation for the $i$-th fluid, $\theta_{i}\equiv i k^{j} v_{j}$ is the divergence
of the $i$-th fluid velocity,  $h = h^{j}_{j}$ is the trace of the metric perturbations $h_{ij}$,
and $\sigma_i$ is the anisotropic stress of the $i$-th fluid. Note that in the following we set
$\sigma_i \equiv 0$ for all $i$, since we assume zero anisotropic stress for all fluids. 
Finally, 
$c_{a,i}^2 = \dot{p}_i/\dot{\rho}_i$ is the adiabatic speed of sound of the $i$-th 
fluid, and it is given by $ c^2_{a,i} =  w_i - \frac{w_i^{\prime}}{3\mathcal{H}(1+w_i)}$
in the case where we set the sound  speed $c^2_{s} = \delta p_i / \delta \rho_i$ to $1$.

\section{Observational Data}
\label{sec-data}

In this section we present the various observational data set that 
are going to be used in order to confront dark energy parametrizations 
with pivoting redshift. In our analysis we incorporate the data  by varying nine
cosmological parameters: the baryon energy density $\Omega_bh^2$, the cold dark matter
energy density $\Omega_ch^2$, the ratio between the sound horizon and the angular
diameter distance at decoupling $\Theta_{s}$, the reionization optical depth $\tau$, 
the spectral index of the scalar perturbations $n_\mathrm{S}$, the amplitude of the
primordial power spectrum $A_\mathrm{S}$, the two parameters of the CPL parametrization
$w_0$ and $w_a$ and the pivot $z_p$. Furthermore, we explore all   parameters
within the range of the conservative flat priors shown in Table~\ref{priors}.
\begin{table}[ht]
\begin{center}
\begin{tabular}{|c|c|}
\hline
Parameter                    & Prior\\
\hline 
$\Omega_{\rm b} h^2$         & $[0.005,0.1]$\\
$\Omega_{\rm c} h^2$                           & $[0.01, 0.99]$\\
$\tau$                       & $[0.01,0.8]$\\
$n_s$                        & $[0.5, 1.5]$\\
$\log[10^{10}A_{s}]$         & $[2.4,4]$\\
$100\theta_{MC}$             & $[0.5,10]$\\ 
$w_0$                        & $[-2, 0]$\\
$w_a$                        & $[-3, 3]$\\
$z_p$                        & $[0, 5]$\\
\hline
\end{tabular}
\end{center}
\caption{The flat priors on the cosmological parameters used in the present analyses. }
\label{priors}
\end{table}

Let us now present in detail the  data sets that we will use.
 
\begin{itemize}

\item Cosmic microwave background (CMB): We constrain the parameters by analyzing the full
range of the 2015 Planck temperature and polarization power spectra 
($2\leq\ell\leq2500$) \cite{Adam:2015rua, Aghanim:2015xee}. This dataset is identified as 
the Planck TTTEEE+lowTEB. At the time of writing only the Planck 2015 likelihood was publicly available,
however we do not expect the conclusions of this paper to change significantly given 
the similarities between Planck 2015 and Planck 2018 results \cite{Akrami:2018vks,planckparams2018}.

\item Baryon acoustic oscillations (BAO): We consider the baryon acoustic oscillations  
as was done in \cite{planckparams2015}.
They are the 6dF Galaxy Survey (6dFGS) measurement at $z_{\emph{\emph{eff}}}=0.106$
\cite{Beutler:2011hx}, the Main Galaxy Sample of Data Release 7 of Sloan Digital Sky Survey (SDSS-MGS) at $z_{\emph{\emph{eff}}}=0.15$ \cite{Ross:2014qpa}, and the CMASS and LOWZ samples from the Data
Release 12 (DR12) of the Baryon Oscillation Spectroscopic Survey (BOSS) at 
$z_{\mathrm{eff}}=0.57$ and at $z_{\mathrm{eff}%
}=0.32$ \cite{Gil-Marin:2015nqa}.

\item Redshift space distortion (RSD): We add two redshift space distortion   data.
In particular, we include the data from CMASS and LowZ galaxy samples. The
CMASS sample consists of 777202 galaxies having the effective redshift of $%
z_{\mathrm{eff}}=0.57$ \cite{Gil-Marin:2016wya}, 
whereas the LOWZ sample consists of 361762 galaxies having an effective redshift of
$z_{\mathrm{eff}}=0.32$~\cite{Gil-Marin:2016wya}.

\item  Weak lensing (WL): We include the cosmic shear data
from the Canada$-$France$-$Hawaii Telescope
Lensing Survey (CFHTLenS) \cite{Heymans:2013fya,Erben:2012zw,Asgari:2016xuw}.

\item Joint light curve analysis (JLA): We consider the joint light curve analysis  
sample  \cite{Betoule:2014frx}  consisting of 740 luminosity distance measurements
of Supernovae Type Ia data in the 
redshift interval $z \in [0.01, 1.30]$.

\item Cosmic chronometers (CC):  We add the thirty measurements of the cosmic chronometers   
in the redshift interval $0< z< 2$. The CC data have 
been summarized in \cite{Moresco:2016mzx}.

\item   We include a Gaussian prior on the Hubble constant value from Riess et al. 
\cite{Riess:2016jrr} (i.e. $H_0=73.24\pm 1.75 \ \rm{km \ s^{-1} \ Mpc^{-1}}$), referred as HST.

\end{itemize}

In order to incorporate statistically the several combinations of datasets and
extract the observational constraints, we   use our modified version of the publicly 
available Monte-Carlo Markov Chain package \texttt{Cosmomc} \cite{Lewis:2002ah}, which  
is an efficient Monte Carlo algorithm equipped with a convergence
diagnostic based on the Gelman and Rubin statistic \cite{Gelman-Rubin}.
It implements an efficient sampling of the posterior distribution using the fast/slow 
parameter decorrelations \cite{Lewis:2013hha} and additionally it includes 
the support for the Planck data release 2015 Likelihood Code \cite{Aghanim:2015xee} \footnote{This code is publicly available at \url{http://cosmologist.info/cosmomc/.}}.

\begin{center}                                                                                                                  
\begin{table*}                                                                                                                   
\begin{tabular}{ccccccccccccccc}                                \hline\hline                                                                                                                    
Parameters & CMB & CBR &~ CBH &~CBRWJCH\\ \hline

$\Omega_c h^2$ & $   0.1191 \pm 0.0014$ & $    0.1192 \pm 0.0013$ & $   0.1194 \pm 0.0013$ & $    0.1183 \pm 0.0012$\\

$\Omega_b h^2$ & $0.02228 \pm 0.00016$ & $    0.02226_{-    0.00016}^{+    0.00014}$ & $    0.02226 \pm 0.00015$ & $    0.02231 \pm  0.00015$ \\

$100\theta_{MC}$ & $1.04080 \pm 0.00033 $ & $    1.04077_{-    0.00031}^{+    0.00032}$ & $    1.04076 \pm 0.00033$ & $    1.04088 \pm   0.00031$ \\

$\tau$ & $ 0.076 \pm 0.018$ & $    0.074 \pm 0.017$ & $    0.078 \pm 0.018$ & $    0.064 \pm 0.017$\\

$n_s$ & $ 0.9665 \pm 0.0045$ & $    0.9661_{-    0.0042}^{+    0.0045}$ & $    0.9658_{-    0.0045}^{+    0.0046}$ 
& $    0.9676 \pm 0.0044$  \\

${\rm{ln}}(10^{10} A_s)$ & $  3.085 \pm 0.034$ & $    3.081 \pm 0.034$ & $    3.089 \pm  0.034$ & $    3.058 \pm  0.033$ \\

$w_0$ & $<-1.3$ & $   -1.33_{-    0.31}^{+    0.35}$ & $   -1.19_{-    0.11}^{+    0.20}$
& $   -1.11_{-    0.11}^{+    0.17}$ \\

$w_a$ & $-0.5_{-    2.0}^{+    1.2}$ & $   -1.21_{-    0.62}^{+    0.62}$ & $   -0.25_{-    0.47}^{+    0.57}$ & $   -0.24_{-    0.33}^{+    0.38}$ \\

$z_p$ & ${\rm unconstrained} $ & $ {\rm unconstrained}$ & $ {\rm unconstrained}$ & $ {\rm unconstrained}$ \\

$\Omega_{m0}$ & $ 0.217_{-    0.076}^{+    0.026} $ & $    0.340 \pm 0.017$ & $    0.291_{-    0.015}^{+    0.014}$ & $ 0.3011_{-    0.0081}^{+    0.0080}$ \\

$\sigma_8$ & $0.96_{-    0.06}^{+    0.11}$ & $    0.804 \pm 0.016$ & $    0.854_{-    0.021}^{+    0.020}$ & $    0.819_{-    0.013}^{+    0.014}$ \\

$H_0$ & $83_{-    8}^{+   14}$ & $   64.7_{-    1.7}^{+ 1.5}$ & $   70.0_{-    1.8}^{+    1.7}$ & $   68.52_{-    0.91}^{+    0.85}$ \\
\hline
$\chi^2_{\mbox{min (best-fit)}}$ & 12960.50 & 12969.720 & 12975.612 & 13723.308\\
\hline\hline                                                                                                                    
\end{tabular}                                                                                                                   
\caption{
Summary of the 68\% CL constraints on  the generalized CPL parametrization 
(\ref{cpl-pivot}), in the case where the pivoting redshift $z_p$ is handled as a free parameter,
using various combinations of the observational 
data sets. Here, CBR = CMB+BAO+RSD, CBH = CMB+BAO+HST, and CBRWJCH = CMB+BAO+RSD+WL+JLA+CC+HST.  }
\label{tab:varying-zp}                                                                                                   
\end{table*}                                                                                                                     
\end{center}

\begin{figure*}[!]
\includegraphics[width=0.65\textwidth]{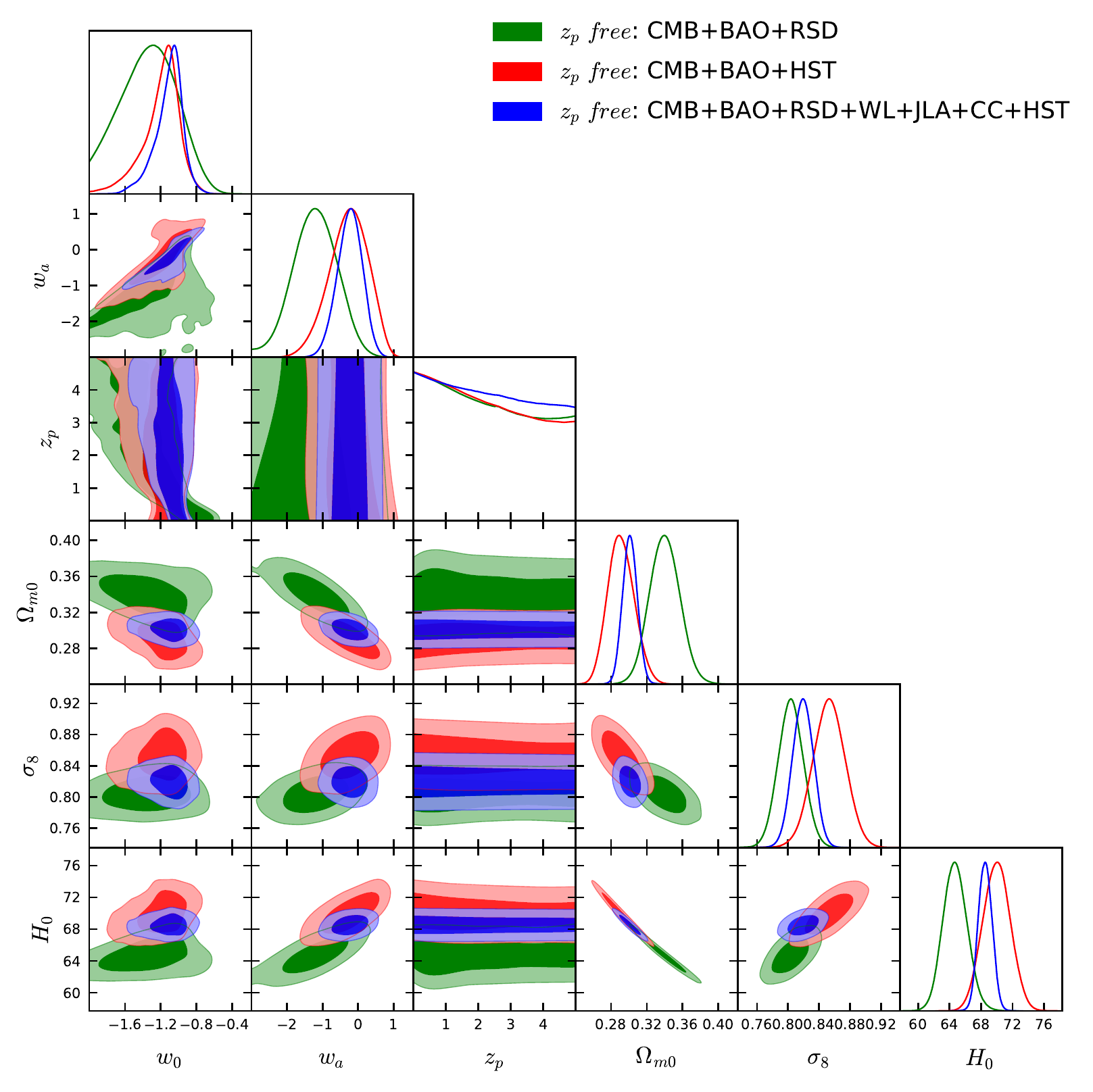}
\caption{{\it{The 68\% and 95\% CL  2-dimensional (2D) contour 
plots for several combinations of various quantities and using
various combinations of the observational data sets, 
for the generalized CPL parametrization (\ref{cpl-pivot}) in the
case where the pivoting redshift $z_p$ is handled as a free parameter, 
and the corresponding 1-dimensional (1D) 
marginalized posterior distributions.}} }
\label{fig-varying-zp}
\end{figure*}

\begin{figure*}[ht]
\includegraphics[width=0.35\textwidth]{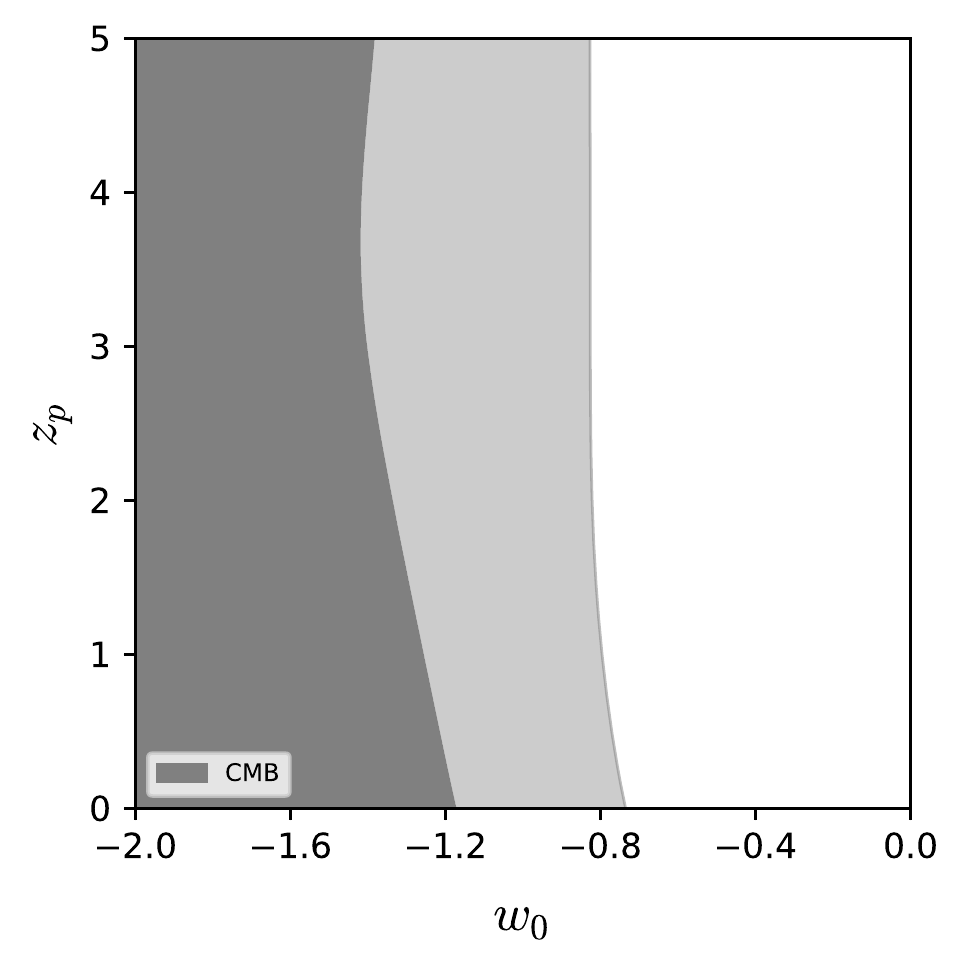}
\includegraphics[width=0.35\textwidth]{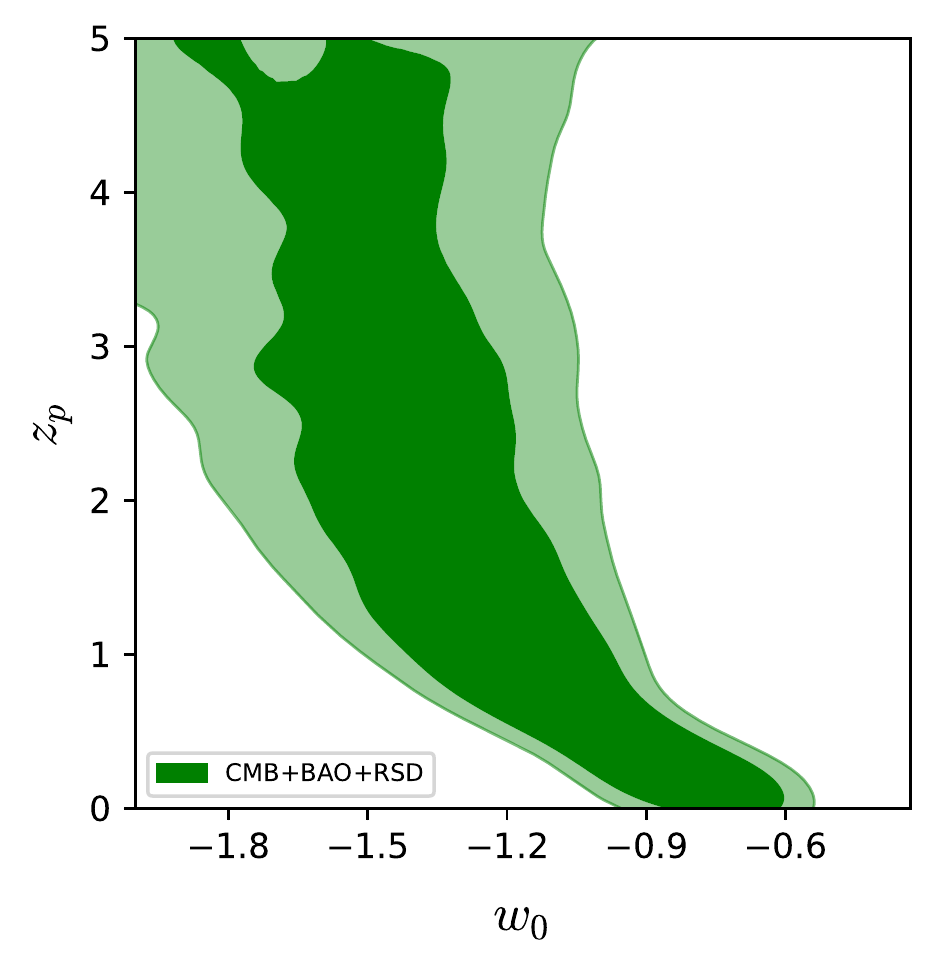}\\
\includegraphics[width=0.35\textwidth]{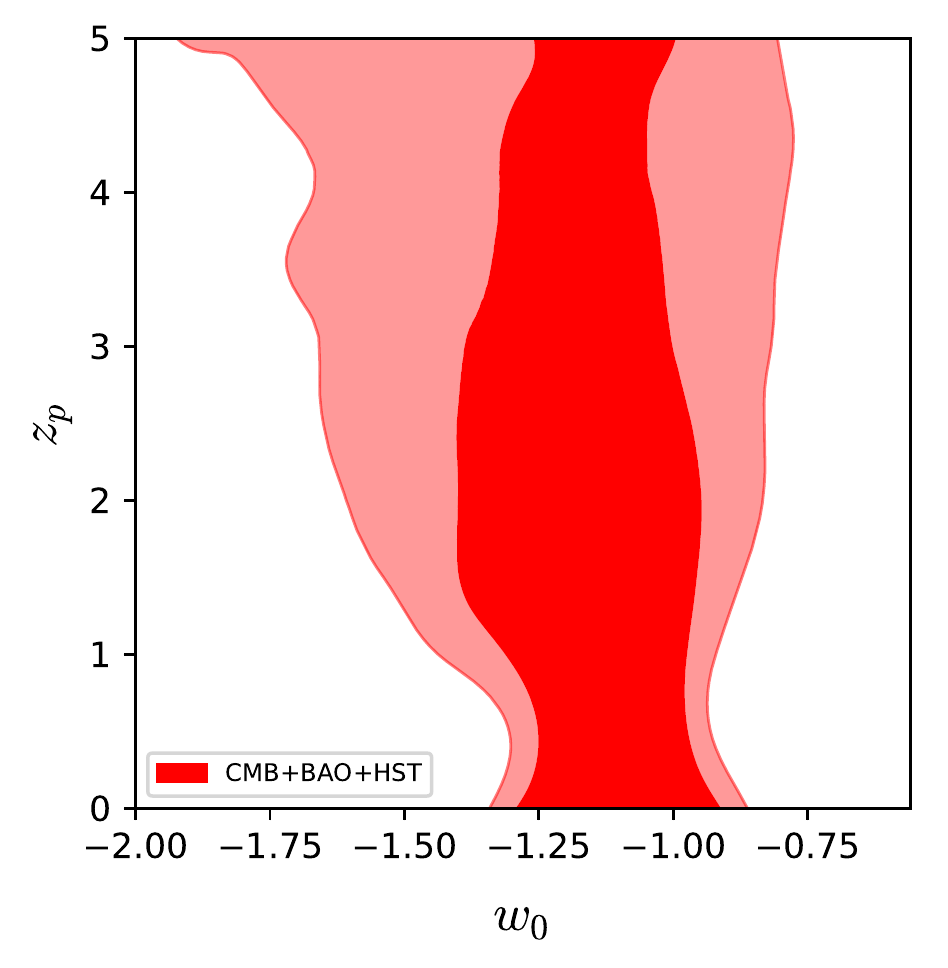}
\includegraphics[width=0.35\textwidth]{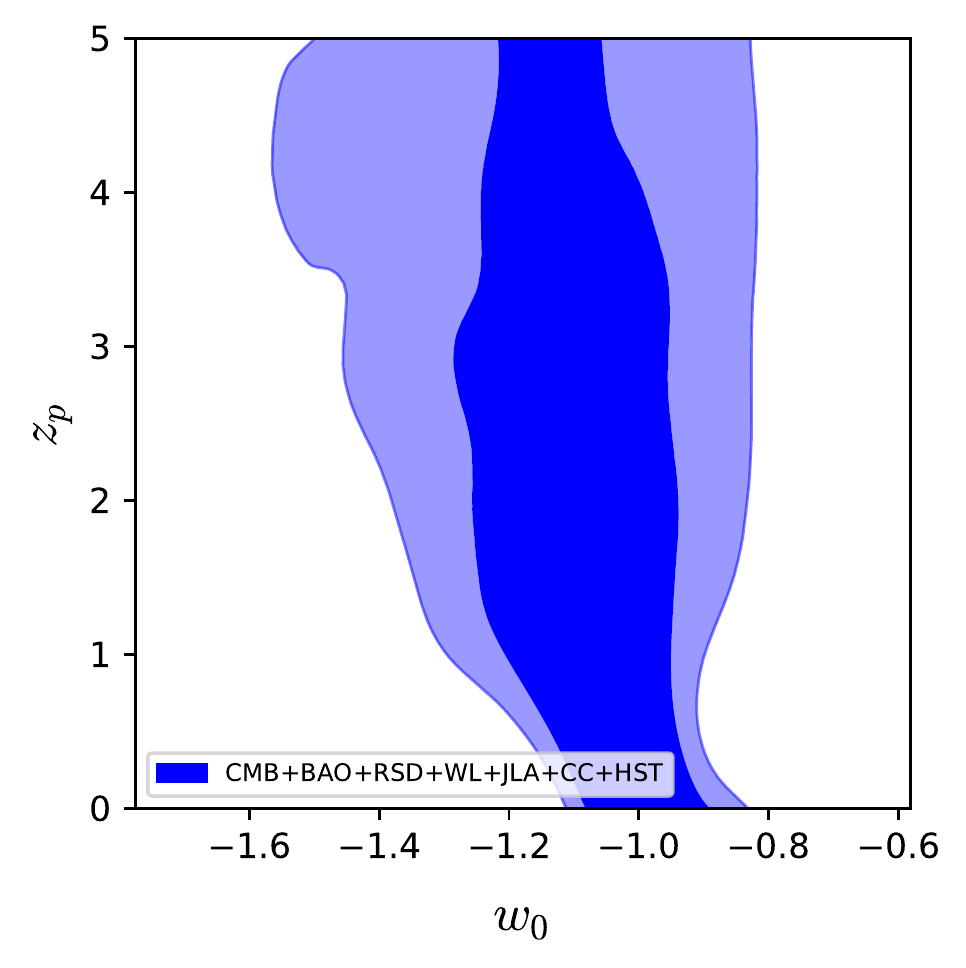}
\caption{{\it{68\% and 95\% CL contour plots in the $w_0-z_p$ plane using various combinations of the observational 
data sets, for the generalized CPL parametrization 
(\ref{cpl-pivot}) in the case where the pivoting redshift $z_p$ is handled as a free parameter.}}}
\label{w0zp}
\end{figure*}

\begin{center}                                                                                                                  
\begin{table*}                                                                                                                   
\begin{tabular}{ccccccc}                                                                                                            
\hline\hline                                                                                                                    
Parameters & CMB & CBR &~ CBH &~CBRWJCH\\ \hline

$\Omega_c h^2$ & $    0.1190_{-    0.0013}^{+    0.0013}$ & $    0.1192_{-    0.0013}^{+    0.0013}$ & $    0.1193_{-    0.0014}^{+    0.0014}$ & $    0.1181_{-    0.0012}^{+    0.0011}$\\

$\Omega_b h^2$ & $    0.02229_{-    0.00017}^{+    0.00015}$ & $    0.02226_{-    0.00015}^{+    0.00015}$ & $    0.02226_{-    0.00017}^{+    0.00015}$ & $    0.02233_{-    0.00015}^{+    0.00016}$ \\

$100\theta_{MC}$ & $    1.04079_{-    0.00033}^{+    0.00033}$ & $    1.04076_{-    0.00031}^{+    0.00031}$ & $    1.04076_{-    0.00033}^{+    0.00032}$ & $    1.04089_{-    0.00031}^{+    0.00031}$ \\

$\tau$ & $    0.077_{-    0.016}^{+    0.016}$ & $    0.074_{-    0.016}^{+    0.016}$ & $    0.079_{-    0.017}^{+    0.018}$ & $    0.066_{-    0.018}^{+    0.017}$\\

$n_s$ & $    0.9668_{-    0.0044}^{+    0.0044}$ & $    0.9662_{-    0.0044}^{+    0.0045}$ & $    0.9660_{-    0.0043}^{+    0.0043}$ 
& $    0.9682_{-    0.0043}^{+    0.0042}$ \\

${\rm{ln}}(10^{10} A_s)$ & $    3.087_{-    0.032}^{+    0.035}$ & $    3.081_{-    0.032}^{+    0.032}$ & $    3.090_{-    0.033}^{+    0.034}$ & $    3.061_{-    0.034}^{+    0.033}$ \\

$w_0$ & $   -1.28_{-    0.44}^{+    0.34}$ & $   -0.68_{-    0.19}^{+    0.24}$ & $   -1.05_{-    0.18}^{+    0.13}$ & $   -1.001_{-    0.078}^{+    0.061}$ \\

$w_a$ & $   -0.9_{-    1.2}^{+    1.4}$ & $   -1.10_{-    0.92}^{+    0.70}$ & $   -0.20_{-  0.41}^{+    0.61}$ 
& $   -0.09_{-  0.20}^{+  0.32}$ \\

$\Omega_{m0}$ & $  0.226_{-    0.084}^{+    0.036}$ & $    0.337_{-    0.020}^{+    0.019}$ & $    0.290_{-    0.017}^{+    0.013}$ & $ 0.3001 \pm 0.0078$ \\

$\sigma_8$ & $    0.95_{-    0.08}^{+    0.12}$ & $    0.805_{-    0.016}^{+    0.016}$ & $    0.854_{-    0.022}^{+    0.022}$ & $    0.819_{-    0.014}^{+    0.014}$ \\

$H_0$ & $   81_{-   10}^{+   14}$ & $   65.0_{-    1.9}^{+    1.7}$ & $   70.1_{-    1.7}^{+    2.0}$ & $   68.58_{-    0.83}^{+    0.85}$ \\
\hline 
$\chi^2_{\mbox{min (best-fit)}}$ & 12961.940 & 12971.976 & 12976.132 & 13724.29\\
\hline\hline                                                    \end{tabular}                                               
\caption{
Summary of the 68\% CL constraints on  the generalized CPL parametrization 
(\ref{cpl-pivot}), in the case where the pivoting redshift is fixed at  $z_p =0.05$,
using various combinations of the observational 
data sets. Here, CBR = CMB+BAO+RSD, CBH = CMB+BAO+HST, and CBRWJCH = CMB+BAO+RSD+WL+JLA+CC+HST. 
}
\label{tab:zp=0.05}                                                                                                   
\end{table*}                                                                                                                     
\end{center}                                                                                                                    
                                                                                                                      
\begin{figure*}[!]
\includegraphics[width=0.65\textwidth]{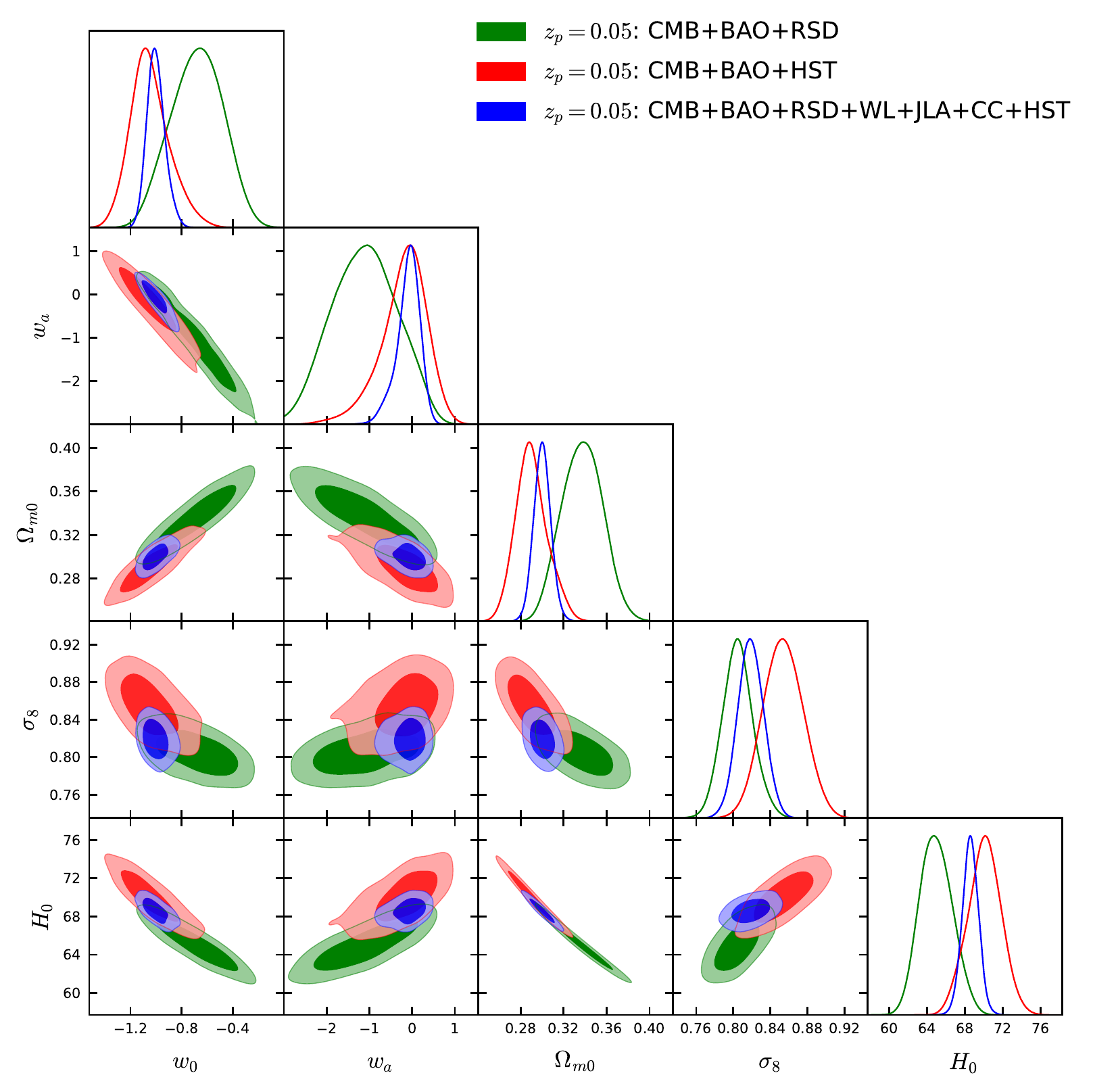}
\caption{{\it{The 68\% and 95\% CL   2D contour plots for several combinations of various quantities and using
various combinations of the observational 
data sets, 
for the generalized CPL parametrization 
(\ref{cpl-pivot}) in the case where the pivoting redshift is fixed at  $z_p =0.05$, and the corresponding  1D
marginalized posterior distributions.}}  }
\label{fig-zp=0.05}
\end{figure*}

\begin{center}                                                                                                                  
\begin{table*}                                                                                                                   
\begin{tabular}{ccccccccccccc}      
\hline\hline                                                    
Parameters & CMB & CBR &~ CBH &~CBRWJCH\\ \hline

$\Omega_c h^2$ & $    0.1191_{-    0.0015}^{+    0.0014}$ & $    0.1191_{-    0.0014}^{+    0.0013}$  & $    0.1193_{-    0.0013}^{+    0.0013}$ & $    0.1181_{-    0.0013}^{+    0.0012}$\\

$\Omega_b h^2$ & $    0.02228_{-    0.00015}^{+    0.00015}$ & $    0.02225_{-    0.00015}^{+    0.00015}$ & $    0.02225_{-    0.00014}^{+    0.00014}$ & $    0.02234_{-    0.00015}^{+    0.00015}$ \\

$100\theta_{MC}$ & $    1.04080_{-    0.00031}^{+    0.00031}$ & $    1.04077_{-    0.00033}^{+    0.00032}$  & $    1.04076_{-    0.00031}^{+    0.00031}$ & $    1.04090_{-    0.00030}^{+    0.00031}$ \\

$\tau$ & $    0.078_{-    0.017}^{+    0.018}$ & $    0.075_{-    0.018}^{+    0.018}$ & $    0.078_{-    0.017}^{+    0.017}$ & $    0.066_{-    0.018}^{+    0.018}$ \\

$n_s$ & $    0.9665_{-    0.0044}^{+    0.0044}$ & $    0.9661_{-    0.0044}^{+    0.0044}$ & $    0.9659_{-    0.0045}^{+    0.0045}$ & $    0.9682_{-    0.0041}^{+    0.0040}$\\

${\rm{ln}}(10^{10} A_s)$ & $  3.088_{-    0.033}^{+    0.035}$ & $    3.082_{-    0.033}^{+    0.037}$ & $    3.090_{-    0.033}^{+    0.034}$ & $    3.061_{-    0.034}^{+    0.034}$ \\

$w_0$ & $   -1.25_{-    0.43}^{+    0.33}$ & $   -0.78_{-    0.16}^{+    0.12}$  & $   -1.05_{-    0.12}^{+    0.09}$ 
& $   -1.010_{-    0.053}^{+    0.045}$ \\

$w_a$ & $   -0.7_{-    1.1}^{+    1.3}$ & $   -1.05_{-    0.59}^{+    0.92}$ & $   -0.27_{-    0.39}^{+    0.64}$  & $   -0.09_{-    0.19}^{+    0.30}$ \\

$\Omega_{m0}$ & $    0.25_{-    0.10}^{+    0.05}$ & $    0.336_{-    0.021}^{+    0.017}$ & $    0.291_{-    0.016}^{+    0.013}$ & $ 0.3001_{-    0.0078}^{+    0.0076}$  \\

$\sigma_8$ & $    0.92_{-    0.10}^{+    0.11}$ & $    0.806_{-    0.016}^{+    0.016}$ & $    0.854_{-    0.021}^{+    0.020}$ & $    0.819_{-    0.014}^{+    0.014}$ \\

$H_0$ & $   78_{-   15}^{+   10}$ & $   65.1_{-    1.7}^{+    1.8}$ & $   70.0\pm  1.7$ & $   68.60_{-    0.90}^{+    0.80}$ \\
\hline 
$\chi^2_{\mbox{min (best-fit)}}$ & 12961.990 & 12970.624 & 12978.076 & 13723.892\\
\hline\hline                                                                                                                    
\end{tabular}                                                                                                                   
\caption{Summary of the 68\% CL constraints on  the generalized CPL parametrization 
(\ref{cpl-pivot}), in the case where the pivoting redshift is fixed at  $z_p =0.15$,
using various combinations of the observational 
data sets. Here, CBR = CMB+BAO+RSD, CBH = CMB+BAO+HST, and CBRWJCH = CMB+BAO+RSD+WL+JLA+CC+HST.  }
\label{tab:zp=0.15}                                             \end{table*}                                               
\end{center}                                                                                                                    
                                                                                                                   
\begin{figure*}[!]
\includegraphics[width=0.65\textwidth]{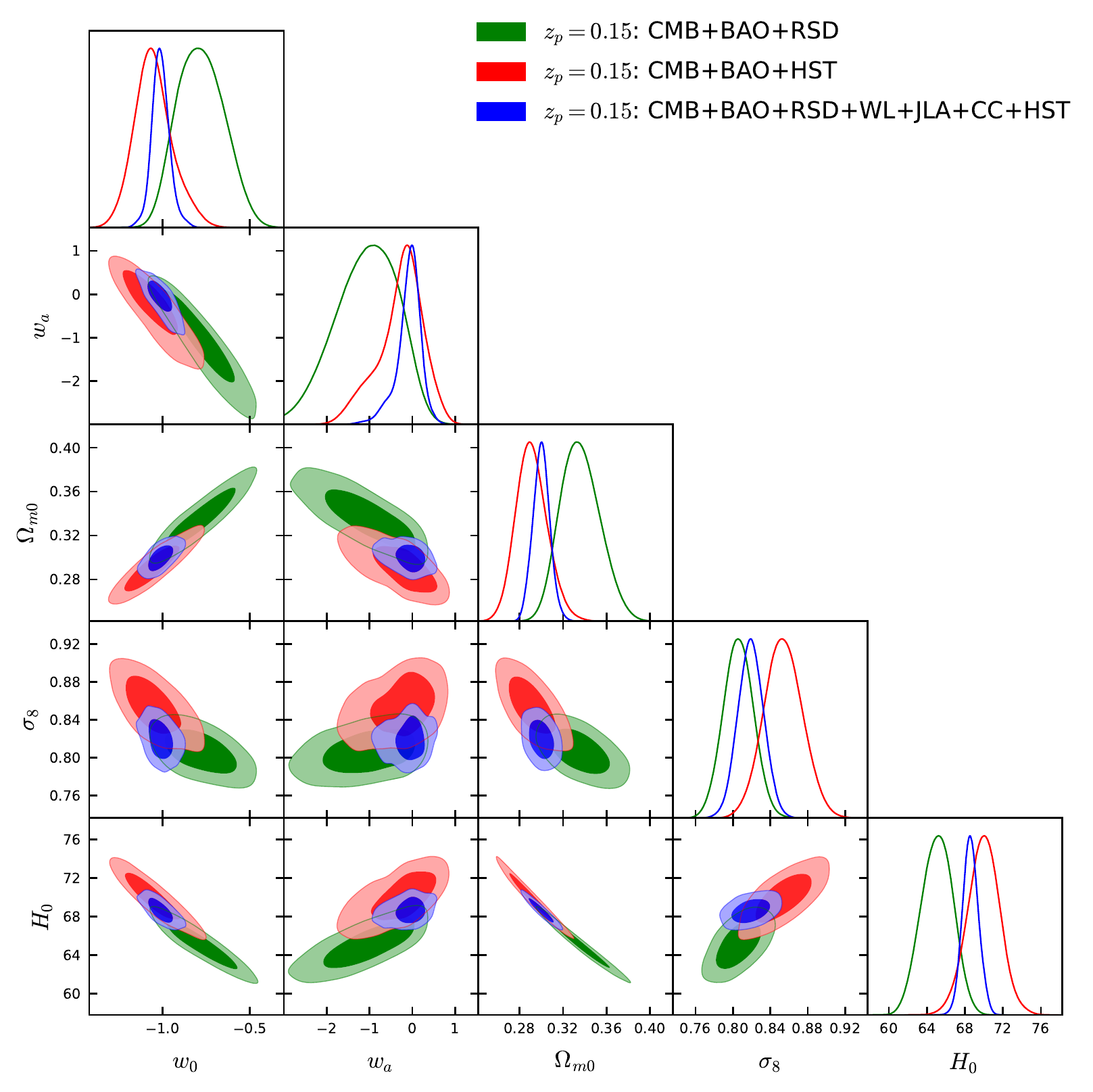}
\caption{{\it{The 68\% and 95\% CL   2D contour plots for several combinations of various quantities and using
various combinations of the observational 
data sets, 
for the generalized CPL parametrization 
(\ref{cpl-pivot}) in the case where the pivoting redshift is fixed at  $z_p =0.15$, and the corresponding  1D
marginalized posterior distributions.}}}
\label{fig-zp=0.15}
\end{figure*}

\section{Observational Constraints}
\label{sec-results}

In this section we provide the observational constraints on the generalized CPL
parametrization with pivoting redshift. We consider two separate cosmological scenarios, 
namely one where the pivoting 
redshift is handled as a free parameter, and one where we fix the pivoting redshift to specific 
values in the region $z_p \in [0, 1]$. Moreover, in order to acquire a complete picture of the behavior
of the scenario, we consider different combinations of the observational
datasets described above.

\subsection{Pivoting redshift as a free parameter}
\label{sec-results-varying-zp}

We desire to impose observational constraints on the   generalized CPL
parametrization (\ref{cpl-pivot}), handling the pivoting redshift $z_p$ as a free parameter.
The results of the analysis  can be seen in Table \ref{tab:varying-zp}, where we display the 68\%  (1$\sigma$) 
confidence level (CL) constraints for various quantities, while the full contour plots 
are presented in Fig.~\ref{fig-varying-zp}.

For the case where only CMB data are used, the Hubble constant value  $H_0$ at present increases and its error 
bars   are strikingly large 
($H_0= 83_{-    8}^{+   14}$ at 68\% CL). Moreover, the present value of the dark-energy equation-of-state parameter is found to lie deeply in the phantom region,
with $ w_0  <-1.3$ at 68\% CL.

When the BAO and RSD data 
are added to CMB (shown in   Table~\ref{tab:varying-zp} as the CBR combination),
$H_0$ decreases ($H_0= 64.7_{- 1.7}^{+ 1.5}$ at 68\% CL) as well as its error bar, while
he matter density increases significantly ($\Omega_{m0} = 0.340\,\pm 0.017$ (at 68\% CL). 
Additionally, in this data combination we obtain   changes into the dark energy constraints.
In particular, although the mean value of the current dark energy equation-of-state parameter 
is in the phantom regime $(w_0= -1.33_{- 0.31}^{+ 0.35})$ at 68\% CL, the  quintessence regime 
 is allowed too, in contrast to the constraints from CMB only. 
 
When the BAO and the HST data are added to CMB (i.e. the combined analysis CMB+BAO+HST called CBH 
in   Table~\ref{tab:varying-zp}), the value of $H_0$ increases again. Concerning $w_0$
we also see  that  a phantom mean value is favored, nevertheless the quintessence regime is allowed within 1$\sigma$.
In addition, the parameter $w_a$ increases significantly  in comparison to its CMB and CMB+BAO+RSD constraints.  However,
as can be seen in Fig.~\ref{fig-varying-zp}, note that the contour plots of the combination of  CMB+BAO+RSD data
(green contours) are in tension with CMB+BAO+HST ones (red contours).

Finally, for the full analysis with all data sets (i.e.  CMB+BAO+RSD+WL+JLA+CC+HST named collectively as CBRWJCH),
summarized in the last column of Table \ref{tab:varying-zp}, we find that the error bars on $H_0$ decrease 
compared to the other three analyses. Furthermore, the value of $w_0$ is in agreement with the cosmological constant
within $1\sigma$, which can also be seen in Fig.~\ref{fig-varying-zp}. 

As we observe, for all combinations of data used,  the pivoting redshift remains unconstrained in the range $[0,5]$, since
$z_p$ remains uncorrelated with most of the cosmological parameters,
with the exception of the current dark-energy equation of state $w_0$.
In order to provide the latter behavior in a clearer way
in Fig.~\ref{w0zp} we present the corresponding contour plots in the $w_0-z_p$ plane. Hence, it is of great importance to
examine the cosmological constraints on the generalized CPL model, handling $z_p$ as a fixed parameter, but still with a value 
different than $z_p=0$ which is its standard CPL value. This is performed in the next subsection.

\subsection{Fixed pivoting redshifts}
\label{sec-results-fixed-zp}

In this subsection we proceed to the investigation of   the generalized CPL
parametrization (\ref{cpl-pivot}),    handling $z_p$ as a fixed parameter in the range 
$[0, 1]$. In particular, we consider six different values of $z_p$, namely 
$z_p = 0.05, 0.15, 0.25, 0.35, 0.50$ and $1$, in order
to examine how the observational constraints will change. 
Moreover, to be uniform we consider the same data combinations  with the previous subsection. 
                         
\begin{center}                                                                                                                  
\begin{table*}                                                                                                                   
\begin{tabular}{cccccccccccccccccc}                                                                                                   
\hline\hline                                                                                                                    
Parameters & CMB & CBR &~ CBH &~CBRWJCH\\ \hline

$\Omega_c h^2$ & $    0.1191_{-    0.0014}^{+    0.0014}$ & $    0.1192_{-    0.0013}^{+    0.0013}$ & $    0.1193_{-    0.0013}^{+    0.0013}$ & $    0.1183_{-    0.0013}^{+    0.0013}$\\

$\Omega_b h^2$ & $    0.02227_{-    0.00017}^{+    0.00015}$ & $    0.02225_{-    0.00016}^{+    0.00015}$ & $    0.02226_{-    0.00015}^{+    0.00015}$  & $    0.02231_{-    0.00015}^{+    0.00015}$ \\

$100\theta_{MC}$ & $    1.04078_{-    0.00035}^{+    0.00034}$ & $    1.04076_{-    0.00032}^{+    0.00032}$ & $    1.04078_{-    0.00033}^{+    0.00032}$ & $    1.04087_{-    0.00032}^{+    0.00032}$ \\

$\tau$ & $    0.078_{-    0.017}^{+    0.017}$ & $    0.075_{-    0.017}^{+    0.018}$ & $    0.077_{-    0.017}^{+    0.017}$ & $    0.064_{-    0.017}^{+    0.017}$ \\

$n_s$ & $    0.9665_{-    0.0046}^{+    0.0045}$ & $    0.9660_{-    0.0043}^{+    0.0043}$ & $    0.9660_{-    0.0045}^{+    0.0044}$ 
& $    0.9675_{-    0.0043}^{+    0.0044}$ \\

${\rm{ln}}(10^{10} A_s)$ & $    3.088_{-    0.034}^{+    0.033}$ & $    3.083_{-    0.033}^{+    0.033}$ & $    3.088_{-    0.033}^{+    0.034}$ & $    3.059_{-    0.034}^{+    0.034}$ \\

$w_0$ & $   -1.40_{-    0.48}^{+    0.28}$ & $   -0.827_{-    0.086}^{+    0.084}$ & $   -1.074_{-    0.081}^{+    0.079}$
& $   -1.007_{-    0.041}^{+    0.041}$ \\

$w_a$ & $   -0.8_{-    0.5}^{+    1.3}$ & $   -1.27_{-    0.64}^{+    0.76}$ & $   -0.24_{-    0.46}^{+    0.66}$ 
& $   -0.24_{-    0.32}^{+    0.38}$ \\

$\Omega_{m0}$ & $    0.229_{-    0.091}^{+    0.043}$ & $    0.341_{-    0.019}^{+    0.017}$ & $    0.291_{-    0.014}^{+    0.014}$ & $ 0.3013_{-    0.0085}^{+    0.0078}$\\

$\sigma_8$ & $    0.95_{-    0.09}^{+    0.12}$ & $    0.804_{-    0.016}^{+    0.016}$ & $    0.853_{-    0.020}^{+    0.020}$ 
& $    0.819_{-    0.014}^{+    0.014}$ \\

$H_0$ & $   81_{-   13}^{+   15}$ &  $   64.6\pm 1.60$  & $   70.0\pm  1.7$  & $   68.50\pm    0.86$ \\

\hline 

$\chi^2_{\mbox{min (best-fit)}}$ & 12959.762 & 12970.200 & 12975.904 & 13722.720\\

\hline\hline                                                                                                                    
\end{tabular}                                                                                                                   
\caption{
Summary of the 68\% CL constraints on  the generalized CPL parametrization  (\ref{cpl-pivot}), in the case where the pivoting redshift is fixed at  $z_p =0.25$,
using various combinations of the observational 
data sets. Here, CBR = CMB+BAO+RSD, CBH = CMB+BAO+HST, and CBRWJCH = CMB+BAO+RSD+WL+JLA+CC+HST. }
\label{tab:zp=0.25}                                                                                                   
\end{table*}                                                                                                                     
\end{center}                                                                                                                    
                                                                                                                
\begin{figure*}[!]
\includegraphics[width=0.65\textwidth]{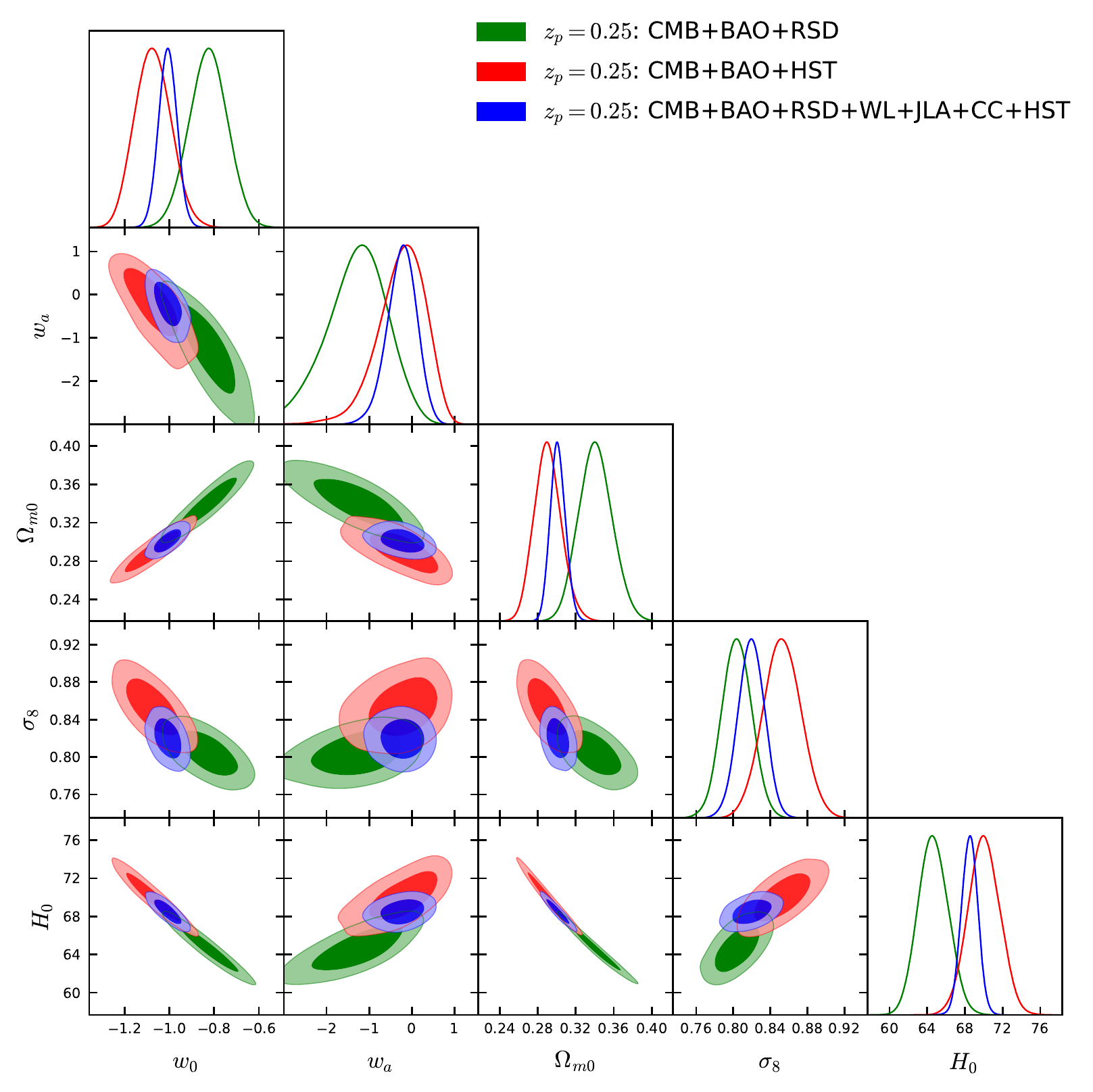}
\caption{{\it{The 68\% and 95\% CL  2D contour plots for several combinations of various quantities and using
various combinations of the observational 
data sets, for the generalized CPL parametrization 
(\ref{cpl-pivot}) in the case where the pivoting redshift is fixed at  $z_p =0.25$, and the corresponding  1D
marginalized posterior distributions. }}}
\label{fig-zp=0.25}
\end{figure*}

\begin{center}                                                                                                                  
\begin{table*}                                                                                                                   
\begin{tabular}{ccccccccccccc}                                                                                                            
\hline\hline                                                                                                                    
Parameters & CMB & CBR &~ CBH &~CBRWJCH\\ \hline

$\Omega_c h^2$ & $    0.1190_{-    0.0014}^{+    0.0014}$ &  $    0.1193_{-    0.0013}^{+    0.0013}$ & $    0.1193_{-    0.0013}^{+    0.0013}$ & $    0.1183_{-    0.0012}^{+    0.0013}$\\

$\Omega_b h^2$ & $    0.02229_{-    0.00016}^{+    0.00016}$ & $    0.02225_{-    0.00015}^{+    0.00015}$ & $    0.02226_{-    0.00016}^{+    0.00014}$ & $    0.02232_{-    0.00015}^{+    0.00015}$ \\

$100\theta_{MC}$ & $    1.04080_{-    0.00032}^{+    0.00033}$ & $    1.04076_{-    0.00032}^{+    0.00031}$ & $    1.04077_{-    0.00032}^{+    0.00032}$ & $    1.04088_{-    0.00031}^{+    0.00032}$ \\

$\tau$ & $    0.076_{-    0.017}^{+    0.017}$ & $    0.074_{-    0.017}^{+    0.017}$ & $    0.078_{-    0.017}^{+    0.017}$ & $    0.065_{-    0.017}^{+    0.018}$ \\

$n_s$ & $    0.9666_{-    0.0049}^{+    0.0046}$ & $    0.9658_{-    0.0044}^{+    0.0044}$ & $    0.9659_{-    0.0044}^{+    0.0043}$ & $    0.9678_{-    0.0044}^{+    0.0043}$ \\

${\rm{ln}}(10^{10} A_s)$ & $    3.085_{-    0.033}^{+    0.033}$ & $    3.081_{-    0.034}^{+    0.033}$ & $    3.089_{-    0.033}^{+    0.034}$ & $    3.061_{-    0.033}^{+    0.034}$ \\

$w_0$ & $   -1.40_{-    0.56}^{+    0.23}$ & $   -0.903_{-    0.055}^{+    0.055}$ & $   -1.083_{-    0.062}^{+    0.061}$ 
& $   -1.022_{-    0.034}^{+    0.033}$ \\

$w_a$ & $   -1.0\pm 1.5$ & $   -1.28_{-    0.70}^{+    0.68}$ & $   -0.30_{-    0.47}^{+    0.64}$ & $   -0.23_{-    0.34}^{+    0.38}$ \\

$\Omega_{m0}$ & $    0.239_{-    0.097}^{+    0.040}$ & $    0.341_{-    0.017}^{+    0.017}$ & $    0.292_{-    0.015}^{+    0.014}$  & $ 0.3007_{-    0.0082}^{+    0.0080}$\\

$\sigma_8$ & $    0.930_{-    0.077}^{+    0.129}$ & $    0.804_{-    0.015}^{+    0.016}$ & $    0.852_{-    0.020}^{+    0.021}$ 
& $    0.820_{-    0.014}^{+    0.014}$\\

$H_0$ & $   80\pm 13$ & $   64.6_{-    1.6}^{+    1.5}$ & $   69.9_{-    1.7}^{+    1.7}$  & $   68.55_{-    0.85}^{+    0.86}$ \\

\hline 

$\chi^2_{\mbox{min (best-fit)}}$ & 12961.850 & 12969.596 & 12976.574 & 13721.112\\

\hline\hline                                                                                                                    
\end{tabular}                                                                                                                   
\caption{Summary of the 68\% CL constraints on  the generalized CPL parametrization 
(\ref{cpl-pivot}), in the case where the pivoting redshift is fixed at  $z_p =0.35$,
using various combinations of the observational 
data sets. Here, CBR = CMB+BAO+RSD, CBH = CMB+BAO+HST, and CBRWJCH = CMB+BAO+RSD+WL+JLA+CC+HST. 
}
\label{tab:zp=0.35}                                                                                                   
\end{table*}                                                                                                                     
\end{center}                                                                                                                    
 
\begin{figure*}[!]
\includegraphics[width=0.65\textwidth]{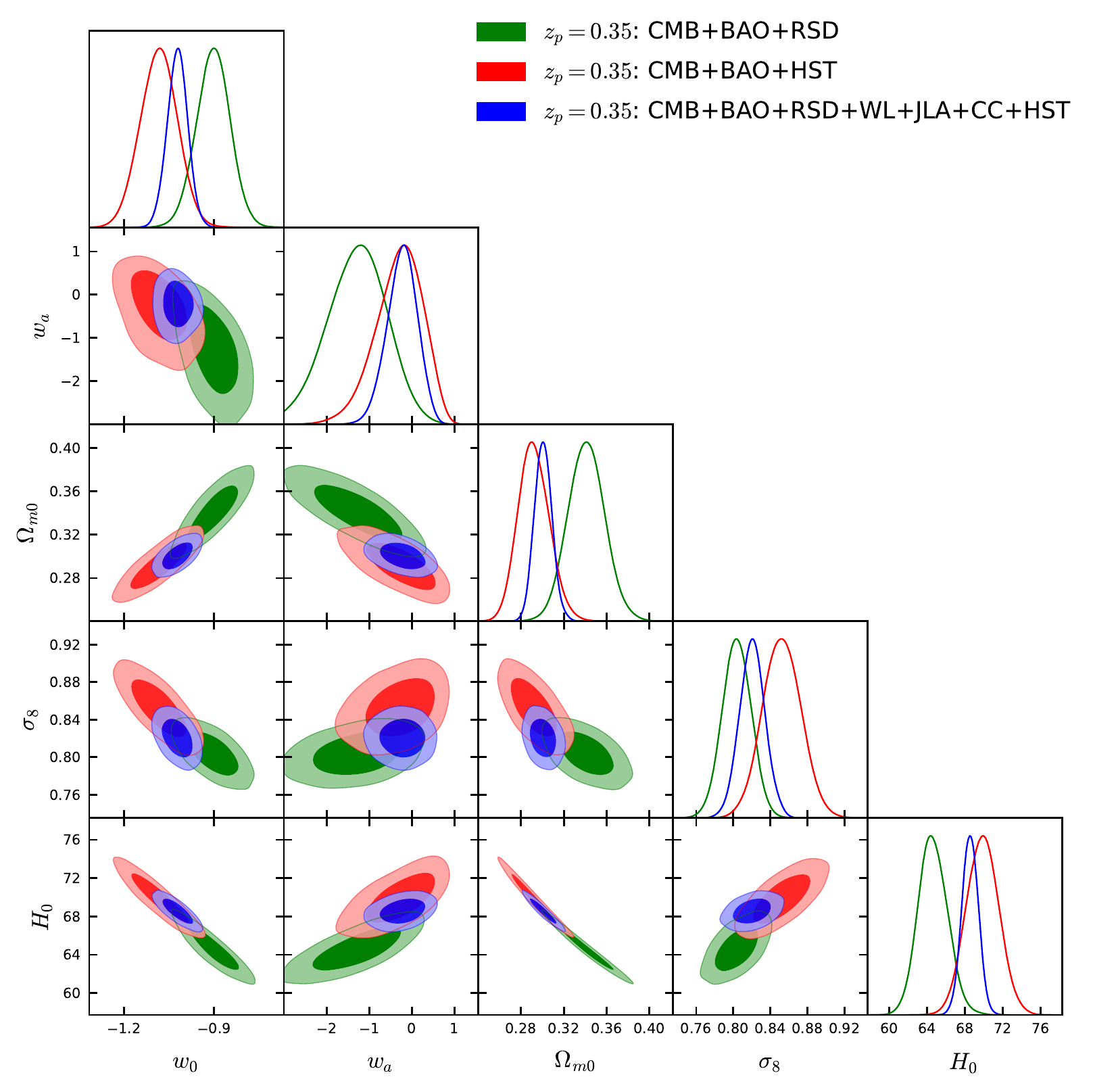}
\caption{{\it{The 68\% and 95\% CL  2D contour plots for several combinations of various quantities and using
various combinations of the observational 
data sets, 
for the generalized CPL parametrization 
(\ref{cpl-pivot}) in the case where the pivoting redshift is fixed at  $z_p =0.35$, and the corresponding  1D
marginalized posterior distributions.}} }
\label{fig-zp=0.35}
\end{figure*}

\subsubsection{Pivoting redshift $z_p = 0.05$}

We begin our analysis  choosing a very small value  $z_p = 0.05$ and we perform the observational fittings
considering several data combinations. The results are summarized in Table \ref{tab:zp=0.05}
and the 68\% and 95\% CL contour plots 
are displayed in Fig. \ref{fig-zp=0.05}.

In the case where we use the CMB data only, the current Hubble constant acquires a large value
$H_0 = 81_{-   10}^{+   14}$ (at 68\% CL), with significantly large error bars.  Concerning the current value of the dark-energy equation-of-state
parameter $w_0$, its mean value lies in the phantom regime  ($w_0 = -1.28_{- 0.44}^{+  0.34}$ at 68\% CL), 
nevertheless it is consistent  with the cosmological constant within one standard deviation.

In the case where we include   BAO and RSD to CMB data (this combination is named as CBR in Table \ref{tab:zp=0.05}) we  obtain lower
$H_0$ values ($H_0 = 65.0_{-1.9}^{+ 1.7}$ at 68\% CL) and its error bars are significantly reduced.
Moreover, $w_0$ lies in the quintessence regime 
in more than 1$\sigma$ ($w_0 = -0.68_{- 0.19}^{+ 0.24}$ at 68\% CL), however  $w_a$  appears to be different from zero at more 
than one standard deviation ($w_a = -1.10_{- 0.92}^{+ 0.70}$  at 68\% CL), due to its strong  anti-correlation with $w_0$. Finally, for this 
data combination  the matter density parameter at present
is rather large ($\Omega_{m0} = 0.337_{- 0.020}^{+ 0.019} $ at 68\% CL) 
compared to the Planck 2015 results \cite{Ade:2015xua}. 

On the other hand, including BAO and HST  to CMB data (this combination is denoted as CBH in Table \ref{tab:zp=0.05})
$H_0$  decreases with respect to the sole CMB case ($H_0= 70.0_{-  1.7}^{+  2.0}$ at 68\% CL), 
but it acquires a higher value with respect to the dataset CBR, in a similar way to what we observed in the free $z_p$ 
analysis presented in section~\ref{sec-results-varying-zp}. Additionally,  $w_0$ is in agreement with the cosmological constant value
at 68\% CL, and similarly, the parameters $w_0$, $w_a$, and $\Omega_{m0}$ are in better agreement with the Planck 2015 findings comparing to
the previous data combination CBR above.

Finally, the full combined analysis CMB+BAO+RSD+WL+JLA+CC+HST produces stronger constraints on the parameters,
and the results are in significantly better agreement with $\Lambda$CDM cosmology.
Concerning the dark-energy equation-of-state parameter we obtain
 $w_0 = -1.001_{-0.078}^{+ 0.061}$ and $w_a =  -0.09_{- 0.20}^{+ 0.32}$  at 68\% CL.

\subsubsection{Pivoting redshift $z_p = 0.15$}

We proceed to the case where the pivot redshift is fixed 
to a slightly larger value, namely $z_p = 0.15$.
The results of the observational confrontation for various datasets
are summarized in Table \ref{tab:zp=0.15},
and the 68\% and 95\% CL contour plots 
are presented in Fig. \ref{fig-zp=0.15}.
 
For the case of CMB data only the increased $z_p$, comparing to the analysis of the previous 
paragraph, leads to smaller $H_0$ values  ($H_0= 78_{- 15}^{+ 10}$ at 68\% CL). On the other hand,
$w_0$ and   $\Omega_{m0}$ increase in comparison to the previous, $z_p = 0.05$, analysis.

In the combined analysis CMB+BAO+RSD we also see that $w_0$ is slightly shifted towards
the cosmological constant comparing to the  $z_p = 0.05$ case. Furthermore, for the last 
two combinations of   datasets, namely, CMB+BAO+HST (CBH   in Table \ref{tab:zp=0.15}) 
and the full CMB+BAO+RSD+WL+JLA+CC+HST (CBRWJCH   in Table \ref{tab:zp=0.15})
we find that the observational constraints are very similar to those obtained for the case with $z_p = 0.05$.

\begin{center}                                                                                                                  
\begin{table*}                                                                                                                   
\begin{tabular}{ccccccccccccccc}                                                                                                            
\hline\hline                                                                                                                    
Parameters & CMB & CBR &~ CBH &~ CBRWJCH\\ \hline

$\Omega_c h^2$ & $    0.1189_{-    0.0013}^{+    0.0013}$ & $    0.1193_{-    0.0013}^{+    0.0013}$ & $    0.1193_{-    0.0013}^{+    0.0013}$ & $    0.1182_{-    0.0012}^{+    0.0012}$ \\

$\Omega_b h^2$ & $    0.02230_{-    0.00016}^{+    0.00015}$ & $    0.02225_{-    0.00015}^{+    0.00015}$ & $    0.02226_{-    0.00015}^{+    0.00015}$ & $    0.02232657_{-    0.00015}^{+    0.00014}$ \\

$100\theta_{MC}$ & $    1.04081_{-    0.00031}^{+    0.00033}$ & $    1.04076_{-    0.00035}^{+    0.00032}$ & $    1.04078_{-    0.00031}^{+    0.00033}$ & $    1.04089_{-    0.00031}^{+    0.00031}$ \\

$\tau$ & $    0.078_{-    0.017}^{+    0.019}$ & $    0.075_{-    0.017}^{+    0.017}$ & $    0.078_{-    0.017}^{+    0.017}$  
& $    0.066_{-    0.016}^{+    0.016}$\\

$n_s$ & $    0.9670_{-    0.0043}^{+    0.0044}$ & $    0.9659_{-    0.0044}^{+    0.0044}$ & $    0.9660_{-    0.0043}^{+    0.0043}$ 
 & $    0.9681_{-    0.0042}^{+    0.0042}$ \\

${\rm{ln}}(10^{10} A_s)$ & $    3.088_{-    0.033}^{+    0.036}$ & $    3.082_{-    0.032}^{+    0.033}$ & $    3.089_{-    0.033}^{+    0.033}$ & $    3.061_{-    0.032}^{+    0.032}$ \\

$w_0$ & $   -1.47_{-    0.40}^{+    0.25}$ & $   -0.996_{-    0.051}^{+    0.061}$ & $   -1.104_{-    0.053}^{+    0.058}$ 
& $   -1.035_{-    0.037}^{+    0.043}$ \\

$w_a$ & $   -0.58_{-    0.35}^{+    0.99}$ & $   -1.29_{-    0.65}^{+    0.77}$ & $   -0.25_{-    0.49}^{+    0.64}$ 
& $   -0.21_{-    0.32}^{+    0.33}$ \\

$\Omega_{m0}$ & $    0.226_{-    0.073}^{+    0.030}$ & $    0.341_{-    0.017}^{+    0.018}$  & $    0.291_{-    0.015}^{+    0.014}$ & $ 0.3008_{-    0.0085}^{+    0.0077}$ \\

$\sigma_8$ & $    0.945_{-    0.071}^{+    0.093}$ & $    0.803_{-    0.016}^{+    0.016}$ & $    0.853_{-    0.020}^{+    0.020}$ 
& $    0.819_{-    0.014}^{+    0.013}$ \\

$H_0$ & $   81_{-    9}^{+   10}$ & $   64.6_{-    1.7}^{+    1.5}$ & $   70.0\pm  1.7$ & $   68.52_{-    0.83}^{+    0.85}$ \\

\hline

$\chi^2_{\mbox{min (best-fit)}}$ & 12960.794 & 12969.644 & 12975.328 & 13723.156\\

\hline\hline                                                                                                                    
\end{tabular}                                                                                                                   
\caption{Summary of the 68\% CL constraints on  the generalized CPL parametrization 
(\ref{cpl-pivot}), in the case where the pivoting redshift is fixed at  $z_p =0.5$,
using various combinations of the observational 
data sets. Here, CBR = CMB+BAO+RSD, CBH = CMB+BAO+HST, and CBRWJCH = CMB+BAO+RSD+WL+JLA+CC+HST. }
\label{tab:zp=0.5}                                                                                                   
\end{table*}                                                                                                                     
\end{center}                                                                                                                    
\begin{figure*}[!]
\includegraphics[width=0.65\textwidth]{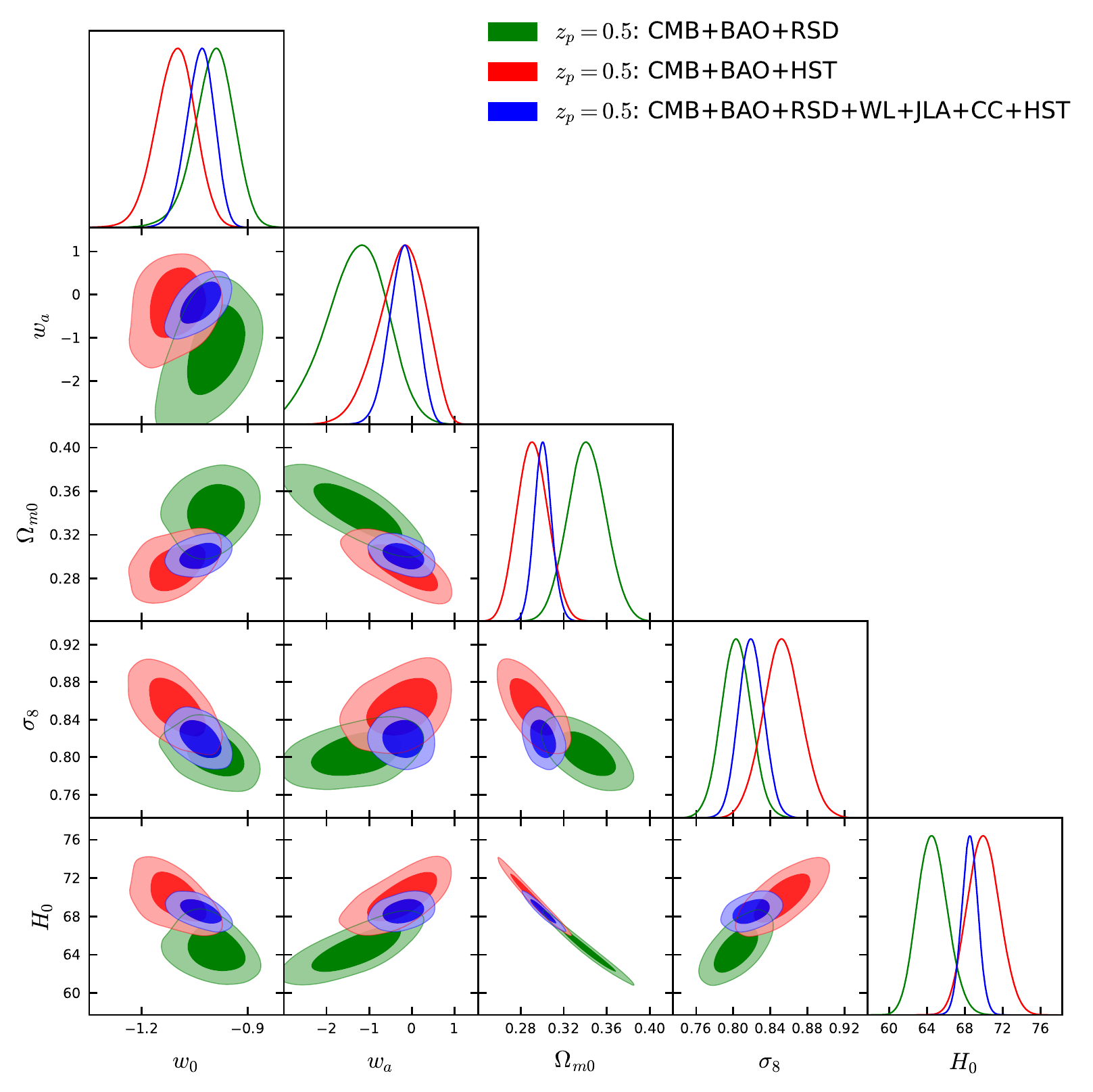}
\caption{{\it{The 68\% and 95\% CL   2D contour plots for several combinations of various quantities and using
various combinations of the observational 
data sets, 
for the generalized CPL parametrization 
(\ref{cpl-pivot}) in the case where the pivoting redshift is fixed at  $z_p =0.5$, and the corresponding  1D
marginalized posterior distributions.}}}
\label{fig-zp=0.5}
\end{figure*}

\subsubsection{Pivoting redshift $z_p = 0.25$}

We fix the pivoting redshift to $z_p = 0.25$ and  in Table \ref{tab:zp=0.25} we summarize the 
fitting results, while in Fig. \ref{fig-zp=0.25} we present the corresponding
68\% and 95\% CL contour plots.

 For the case of CMB data only, $H_0$  is very similar to that obtained for the analysis with $z_p = 0.05$. 
 However, $w_0 < -1$ at more than 68\% CL, while in the previous analyses with  $z_p = 0.05$ and $z_p = 0.15$ we had found $w_0 > -1$ at 1$\sigma$. Additionally, concerning  $w_a$ we observe that its mean value lies  in the middle between the value obtained for $z_p = 0.05$ and $z_p = 0.15$.

 When we add external data sets to CMB we find significant improvements in the
 estimations of the Hubble parameter, and its error bars are one order of magnitude smaller.
 The pattern of the analysis for the combined dataset CMB+BAO+RSD remains the same as
 the previous two analyses with the pivoting redshifts $z_p = 0.05$ and $z_p =0.15$. 
 In fact,  $w_0$ is in the quintessential regime at more than one standard deviation,
 and the matter density parameter shifts towards higher values.
 A significant improvement appears when we consider the two combinations of data sets, 
 namely, CMB+BAO+HST and  CMB+BAO+RSD+WL+JLA+CC+HST, where the various quantities exhibit 
 similar trends   with the previous two fixed pivoting redshifts.
 We mention that the two key parameters of the dark energy parametrization,
 namely $w_0$ and $w_a$, are now in perfect agreement  with the cosmological constant.

\begin{center}                                                                                                                  
\begin{table*}                                                                                                                   
\begin{tabular}{ccccccccccccccc}                                                                                                            
\hline\hline                                                                                                                    
Parameters & CMB & CBR &~ CBH &~ CBRWJCHST \\ \hline

$\Omega_c h^2$ & $    0.1192_{-    0.0014}^{+    0.0014}$ & $    0.1189_{-    0.0014}^{+    0.0014}$ & $    0.1192_{-    0.0014}^{+    0.0013}$ & $    0.1181_{-    0.0012}^{+    0.0012}$ \\

$\Omega_b h^2$ & $    0.02227_{-    0.00016}^{+    0.00016}$ & $    0.02228_{-    0.00016}^{+    0.00015}$ & $    0.02227_{-    0.00015}^{+    0.00015}$ & $    0.02233_{-    0.00014}^{+    0.00014}$ \\

$100\theta_{MC}$ & $    1.04079_{-    0.00033}^{+    0.00032}$ & $    1.04082_{-    0.00033}^{+    0.00033}$ & $    1.04078_{-    0.00031}^{+    0.00031}$ & $    1.04089_{-    0.00031}^{+    0.00032}$ \\

$\tau$ & $    0.077_{-    0.017}^{+    0.017}$ & $    0.076_{-    0.018}^{+    0.018}$ & $    0.079_{-    0.017}^{+    0.017}$ 
& $    0.066_{-    0.017}^{+    0.017}$ \\

$n_s$ & $    0.9663_{-    0.0045}^{+    0.0049}$ & $    0.9669_{- 0.0044}^{+  0.0045}$ & $    0.9661_{-    0.0044}^{+    0.0044}$  
& $    0.9682_{-    0.0043}^{+    0.0043}$ \\

${\rm{ln}}(10^{10} A_s)$ & $    3.087_{-    0.032}^{+    0.033}$ & $    3.083_{-    0.034}^{+    0.035}$ & $    3.091_{-    0.033}^{+    0.035}$ & $    3.062_{-    0.033}^{+    0.033}$ \\

$w_0$ & $ -1.44_{-    0.42}^{+    0.28}$ & $   -1.13_{-    0.11}^{+  0.21}$ & $   -1.12_{-    0.08}^{+    0.13}$ & $   -1.047_{-    0.053}^{+    0.093}$ \\

$w_a$ & $   -0.6_{-    0.4}^{+    1.0}$ & $   -0.88_{-    0.64}^{+    0.98}$ & $   -0.12_{-    0.35}^{+    0.61}$ & $   -0.11_{-    0.22}^{+    0.37}$ \\

$\Omega_{m0}$ & $    0.253_{-    0.094}^{+    0.049}$ & $    0.333_{-    0.022}^{+    0.017}$ & $    0.289_{-    0.015}^{+    0.012}$ & $ 0.3001_{-    0.0083}^{+    0.0082}$ \\

$\sigma_8$ & $    0.91\pm 0.10$ & 
$ 0.806_{-    0.016}^{+  0.016}$ & $    0.854_{-    0.020}^{+    0.020}$ 
& $    0.819_{-    0.014}^{+    0.013}$ \\

$H_0$ & $   77_{-   13}^{+    10}$ & $   65.3 \pm 1.8$ & 
$ 70.2_{-1.6}^{+    1.8}$ &  $  68.59_{-    0.95}^{+    0.86}$ \\

\hline 

$\chi^2_{\mbox{min (best-fit)}}$ & 12960.778 & 12969.896 & 12975.270 & 13721.972\\

\hline\hline                                                                                                                    
\end{tabular}                                                                                                                   
\caption{Summary of the 68\% CL constraints on  the generalized CPL parametrization 
(\ref{cpl-pivot}), in the case where the pivoting redshift is fixed at  $z_p =1$,
using various combinations of the observational 
data sets. Here, CBR = CMB+BAO+RSD, CBH = CMB+BAO+HST, and CBRWJCH = CMB+BAO+RSD+WL+JLA+CC+HST. }
\label{tab:zp=1}                                                                                                   
\end{table*}                                                                                                                     
\end{center}                                                                                                                    
                                                                                                        
\begin{figure*}[!]
\includegraphics[width=0.65\textwidth]{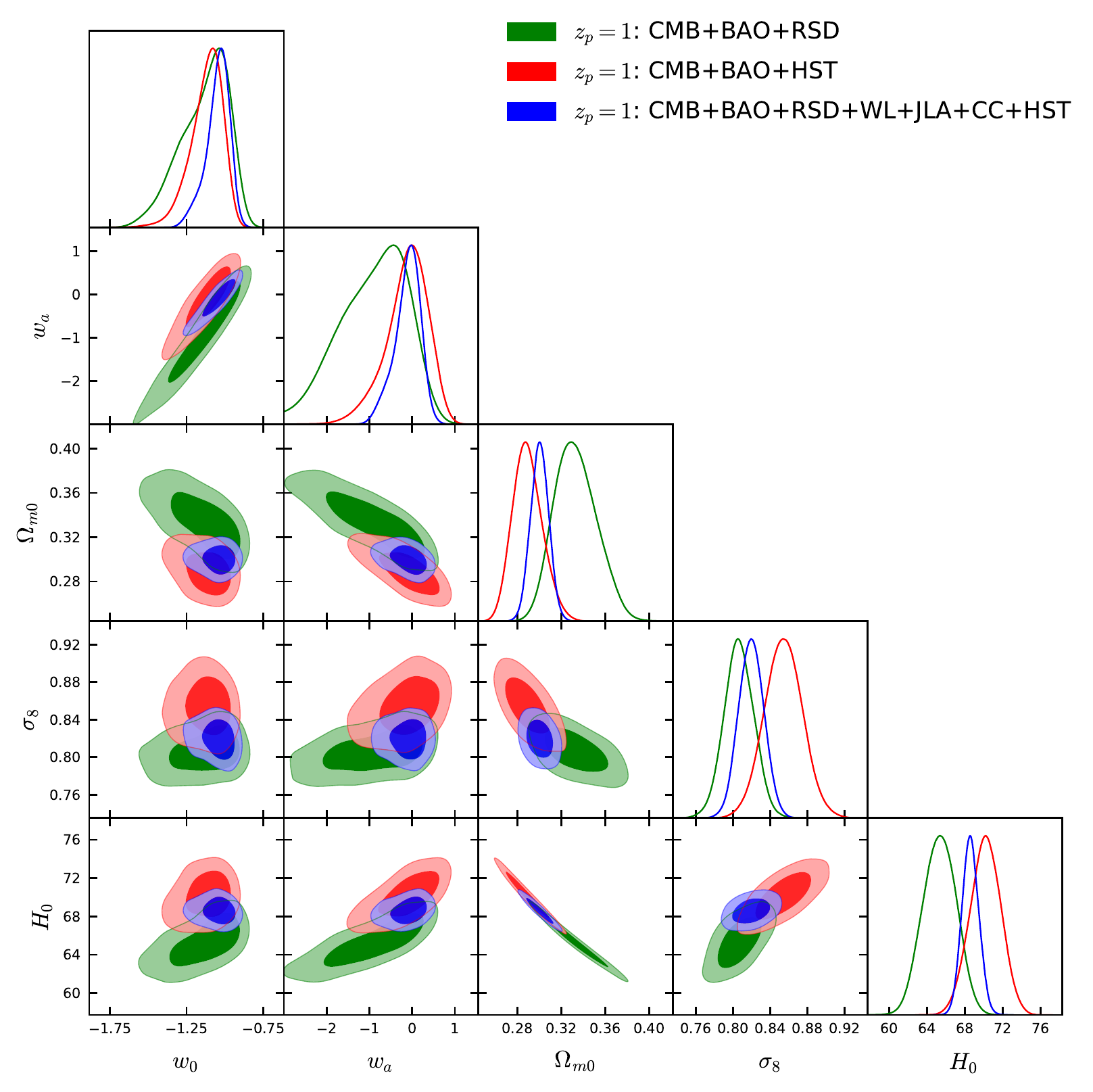}
\caption{{\it{The 68\% and 95\% CL   2D contour plots for several combinations of various quantities and using
various combinations of the observational 
data sets, 
for the generalized CPL parametrization 
(\ref{cpl-pivot}) in the case where the pivoting redshift is fixed at  $z_p =1$, and the corresponding  1D
marginalized posterior distributions.}} }
\label{fig-zp=1}
\end{figure*}

\begin{center}                                                  
\begin{table*}                                                  
\begin{tabular}{cccccccccccccccccc}            
\hline\hline                                                                                                                    
Parameters & CMB & CBR &~ CBH &~ CBRWJCH \\ \hline

$\Omega_c h^2$ & $    0.1191\pm  0.0014$ & $    0.1191_{-    0.0013}^{+    0.0014}$ & $    0.1192_{-    0.0012}^{+    0.0013}$ & $    0.1182\pm  0.0012$\\

$\Omega_b h^2$ & $    0.02228_{-    0.00015}^{+    0.00016}$ & $    0.02226\pm 0.00015$ & $    0.02227\pm    0.00015$ & $    0.02233\pm   0.00014$\\

$100\theta_{MC}$ & $ 1.04078_{-    0.00034}^{+    0.00031}$ & 
$    1.04079_{-    0.00033}^{+    0.00032}$ & $    1.04079_{-    0.00031}^{+    0.00034}$ &  $    1.04090_{-    0.00033}^{+    0.00031}$ \\

$\tau$ & $0.075\pm    0.017$ & $    0.074\pm  0.018$ & $    0.079\pm 0.017$ & $    0.065\pm  0.016$ \\

$n_s$ & $0.9664\pm    0.0045$ & $    0.9662_{-    0.0045}^{+    0.0044}$ & $    0.9664_{-    0.0049}^{+    0.0043}$ & $    0.9679\pm  0.0043$ \\

${\rm{ln}}(10^{10} A_s)$ & $ 3.084\pm   0.033$ & 
$    3.080_{-    0.034}^{+    0.035}$ & $    3.090_{-    0.033}^{+    0.034}$ & $    3.060\pm   0.032$ \\

$w_0$ & $  -1.31_{-    0.56}^{+    0.43}$ &  $   -0.63_{-    0.30}^{+    0.20}$ & $   -1.04\pm 0.17$ & $   -0.980_{-    0.085}^{+    0.067}$ \\
$w_a$ & $   -0.9_{-    1.7}^{+    1.0}$ & $   -1.1_{-    0.6}^{+    1.0}$ & $   -0.18_{-    0.50}^{+    0.56}$ & $   -0.16_{-    0.19}^{+    0.31}$ \\

$\Omega_{m0}$ & $   0.213_{-    0.077}^{+    0.026}$ & $    0.336_{-    0.021}^{+    0.018}$ & $    0.290_{-    0.016}^{+    0.014}$ & $ 0.300\pm    0.008$  \\

$\sigma_8$ & $0.968_{-    0.063}^{+    0.118}$ & $    0.805_{-    0.015}^{+    0.015}$  & $    0.853_{-    0.020}^{+    0.020}$ & $    0.819_{-    0.013}^{+    0.014}$ \\

$H_0$ & $84_{-    8}^{+   15}$ & $   65.0_{-    1.7}^{+    1.8}$ & $   70.1\pm 1.7$ & $   68.59_{-    0.80}^{+    0.80}$ \\

\hline 
$\chi^2_{\mbox{min (best-fit)}}$ & 12958.172 & 12968.590 & 12976.962 & 13723.702\\

\hline\hline                                                              
\end{tabular} 
\caption{Summary of the 68\% CL constraints on  the standard CPL parametrization 
(\ref{cpl}), namely without any pivoting redshift ($z_p =0$),
using various combinations of the observational 
data sets. Here, CBR = CMB+BAO+RSD, CBH = CMB+BAO+HST, and CBRWJCH = CMB+BAO+RSD+WL+JLA+CC+HST.  }
\label{tab:cpl-no-pivot}                  
\end{table*}  
\end{center}

\begingroup                                          
\squeezetable                                                                                                  
\begin{center}                                                                                                                  
\begin{table*}                                                                                                                   
\begin{tabular}{ccccccccccccc}                                                                                                       
\hline\hline                                
Datasets &      
Parameter & $z_p$   free & $z_p =0$ (CPL) & $z_p =0.05$ & $z_p =0.15$ & $z_p =0.25$ & $z_p =0.35$ & $z_p =0.5$ & $z_p =1$ \\ \hline

CBR&
$w_0$ & $   -1.33_{-    0.31}^{+    0.35}$ & $   -0.63_{-    0.30}^{+    0.20}$ & $   -0.68_{-    0.19}^{+    0.24}$ & $   -0.78_{-    0.16}^{+    0.12}$ & $   -0.827_{-    0.086}^{+    0.084}$ & $   -0.903_{-    0.055}^{+    0.055}$ & $   -0.996_{-    0.051}^{+    0.061}$ & $   -1.13_{-    0.11}^{+  0.21}$ \\

CBR&
$w_a$ & $   -1.21_{-    0.62}^{+    0.62}$ & $   -1.1_{-    0.6}^{+    1.0}$ & $   -1.10_{-    0.92}^{+    0.70}$ & $   -1.05_{-    0.59}^{+    0.92}$ & $   -1.27_{-    0.64}^{+    0.76}$ & $   -1.28_{-    0.70}^{+    0.68}$ & $   -1.29_{-    0.65}^{+    0.77}$ & $   -0.88_{-    0.64}^{+    0.98}$ \\

\hline

CBRWJCH&
$w_0$ &  $   -1.11_{-    0.11}^{+    0.17}$  & $   -0.980_{-    0.085}^{+    0.067}$ &  $   -1.001_{-    0.078}^{+    0.061}$  & $   -1.010_{-    0.053}^{+    0.045}$ & $   -1.007_{-    0.041}^{+    0.041}$ & $   -1.022_{-    0.034}^{+    0.033}$ & $   -1.035_{-    0.037}^{+    0.043}$ & $   -1.047_{-    0.053}^{+    0.093}$  \\

CBRWJCH&
$w_a$ & $   -0.24_{-    0.33}^{+    0.38}$ & $   -0.16_{-    0.19}^{+    0.31}$  & $   -0.09_{-  0.20}^{+  0.32}$ & $   -0.09_{-    0.19}^{+    0.30}$ & $   -0.24_{-    0.32}^{+    0.38}$ & $   -0.23_{-    0.34}^{+    0.38}$ & $   -0.21_{-    0.32}^{+    0.33}$ & $   -0.11_{-    0.22}^{+    0.37}$ \\

\hline 
\end{tabular}
\caption{Summary of the constraints on the dark energy parametrization (\ref{cpl-pivot}), for various values of the pivot redshift $z_p$ (free/fixed), using the observational data CBR =CMB+BAO+RSD (upper half of the table) and CBRWJCH=CMB+BAO+RSD+WL+JLA+CC+HST (lower half of the table).}
\label{tab:comparison}                  
\end{table*}                
\end{center}   
\endgroup

\subsubsection{Pivoting redshift $z_p = 0.35$}

We now consider a slightly higher pivot redshift value, namely $z_p= 0.35$, 
and we summarize the fitting results  in Table \ref{tab:zp=0.35}, 
while in Fig. \ref{fig-zp=0.35} we show the corresponding
68\% and 95\% CL contour plots.

The observational pattern for this parametrization is the same as the previous cases.
We find that the CMB data only constrain $w_0 < -1$ at more than 68\% CL,   and we recover 
the cosmological constant scenario as soon as we add more external datasets to CMB. 
In this case, it appears an indication for $w_0 < -1$ at one standard deviation 
also for the CMB+BAO+HST case. Moreover, the results with the full combination of
datasets are very stable and robust towards changing the fixed pivot redshift.

\subsubsection{Pivoting redshift $z_p = 0.5$}

For the case $z_p= 0.50$, we summarize the fitting results  in Table \ref{tab:zp=0.5} and in Fig. \ref{fig-zp=0.5} we present
the corresponding 68\% and 95\% CL contour plots.
As we see, the constraints on the parameters are very similar to the case  $z_p = 0.35$. 
In particular, we have the preference for a phantom regime at one standard deviation for the CMB  and the CMB+BAO+HST cases, while on the other hand for the combinations CMB+BAO+RSD and CMB+BAO+RSD+WL+JLA+CC+HST the parametrization is in agreement with the cosmological constant.

\subsubsection{Pivoting redshift $z_p = 1$}

Finally, we consider the last fixed pivoting redshift in this series, namely $z_p=  1$. 
In Table \ref{tab:zp=1} we summarize the fitting results while in Fig. \ref{fig-zp=1} we depict
the corresponding 68\% and 95\% CL contour plots.
The overall results are similar with respect to the previous fixed pivot redshift cases,
however one can clearly see that for CMB+BAO+RSD data sets $w_0$ has a mean value
in the phantom regime ($w_0 = -1.13_{- 0.11}^{+  0.21}$ at 68\% CL).
Therefore, with the increment of the pivoting redshift we find a successive change 
in the estimation of $w_0$. One can further notice that for the CMB data only $w_0<-1$ is always favored, but all the other combinations of data recover the cosmological constant within 68\% CL.

\begin{figure*}[!]
\includegraphics[width=0.715\textwidth]{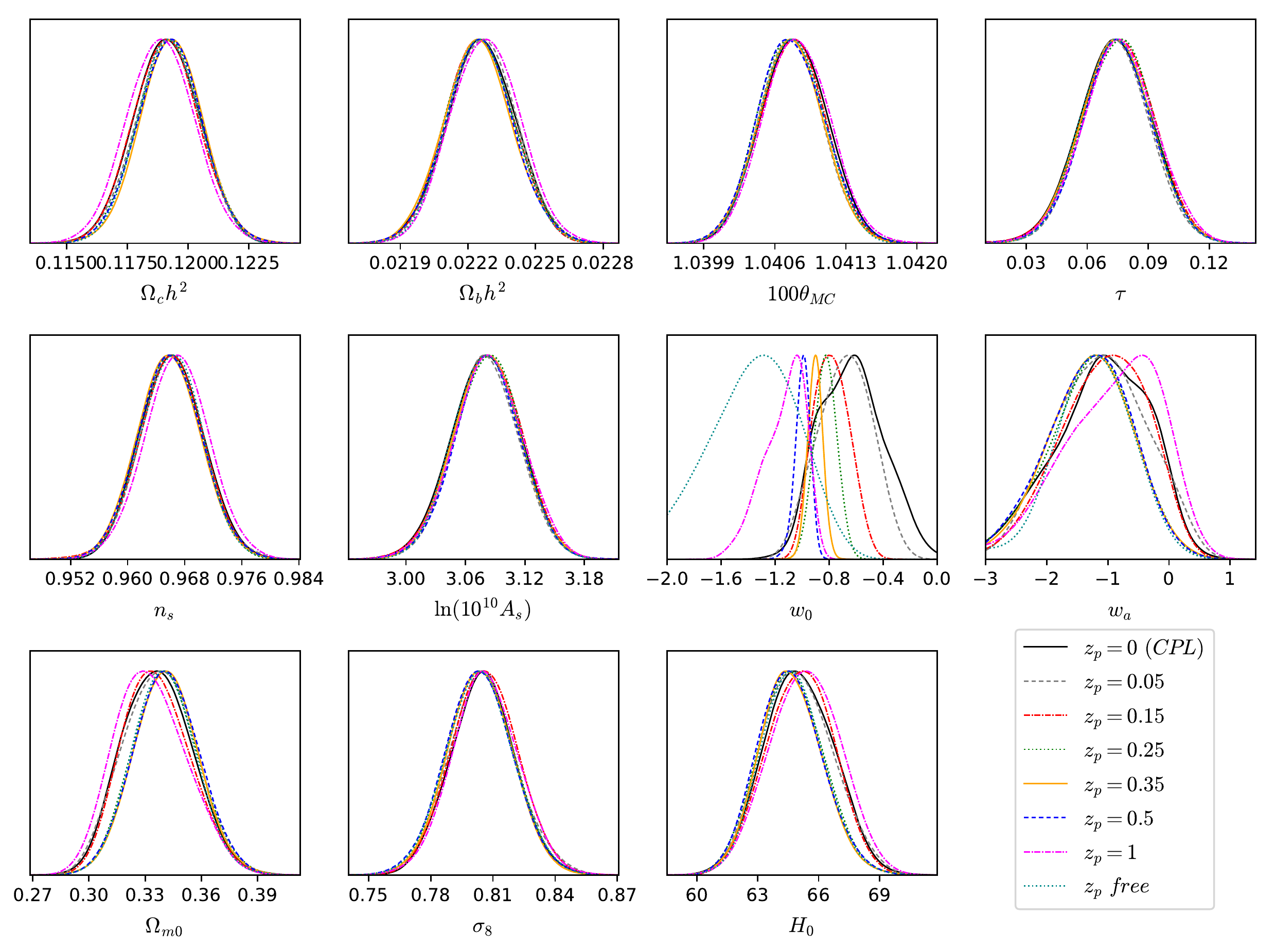}\\
\includegraphics[width=0.715\textwidth]{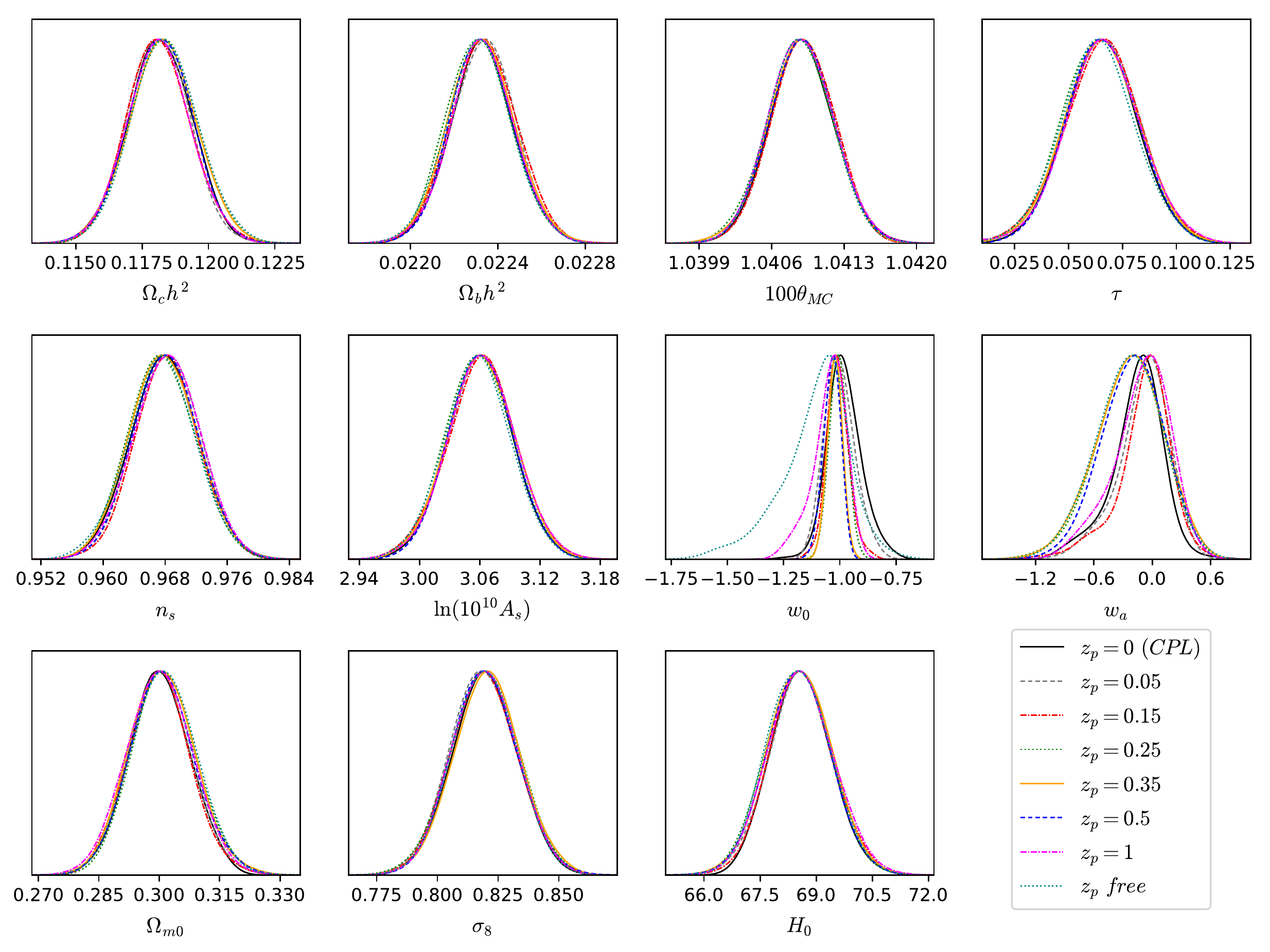}
\caption{{\it{One-dimensional marginalized posterior distributions of various  parameters of the generalized CPL parametrization, for free and fixed pivoting redshift $z_p$, for the combined analyses CMB+BAO+RSD (upper three rows) and  CMB+BAO+RSD+WL+JLA+CC+HST (lower three rows).
One can observe that the various parameters behave similarly, apart from $w_0$ and $w_a$.}}}
\label{fig:comparison1}
\end{figure*}

\begin{figure*}[!]
\includegraphics[width=0.49\textwidth]{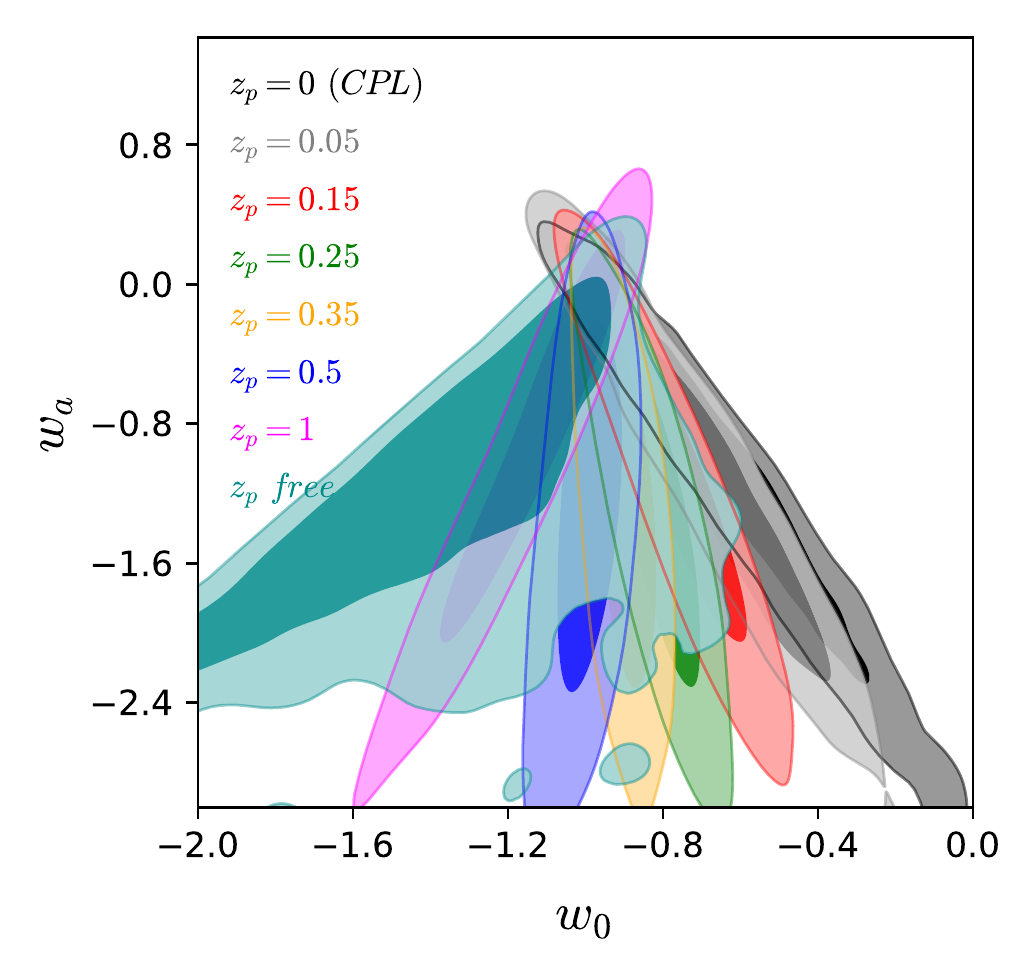}
\includegraphics[width=0.49\textwidth]{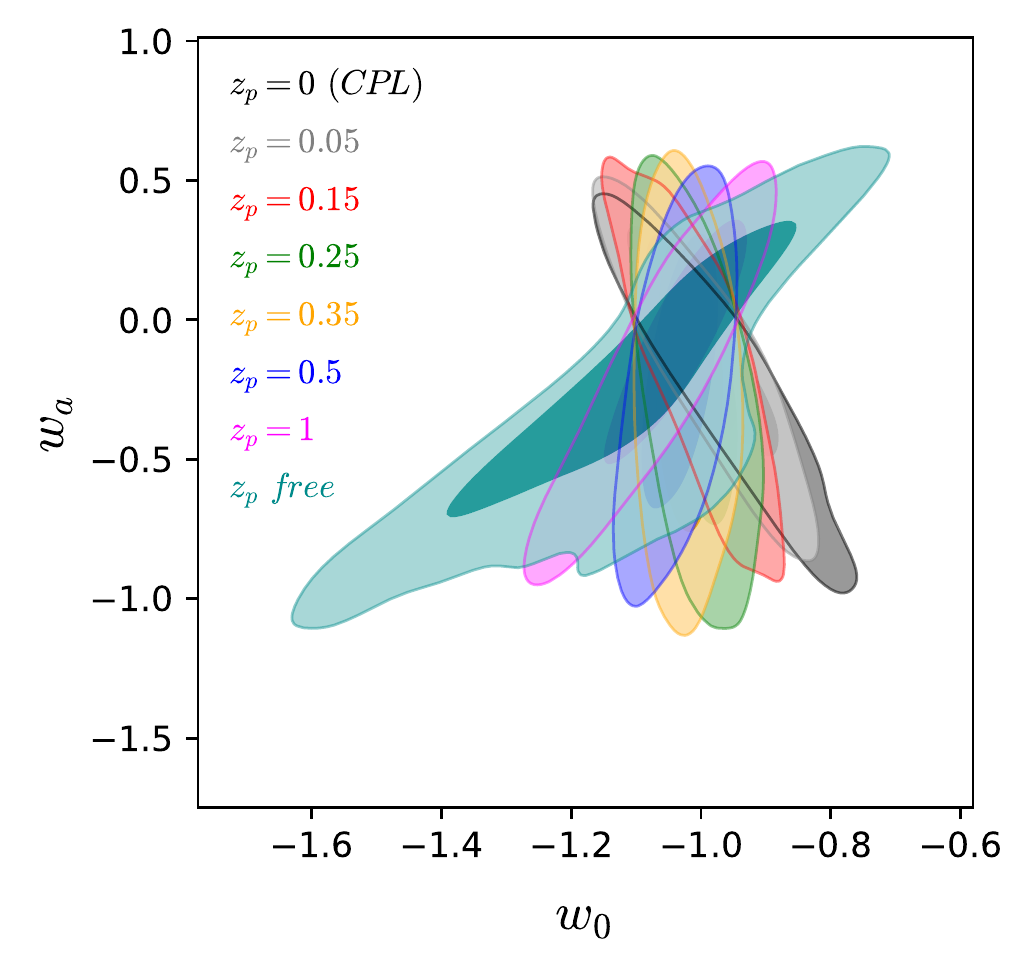}
\caption{{\it{The 68\% and 95\% CL  contour plots
in the $w_0-w_a$ plane,   
for the generalized CPL parametrization 
(\ref{cpl-pivot}) for various pivoting redshifts $z_p$,
and for the data combinations CMB+BAO+RSD (left graph) 
and CMB+BAO+RSD+WL+JLA+CC+HST (right graph). }}}
\label{fig:comparison-w0-wa}
\end{figure*}  

\begin{figure*}[!]
\includegraphics[width=1.0\textwidth]{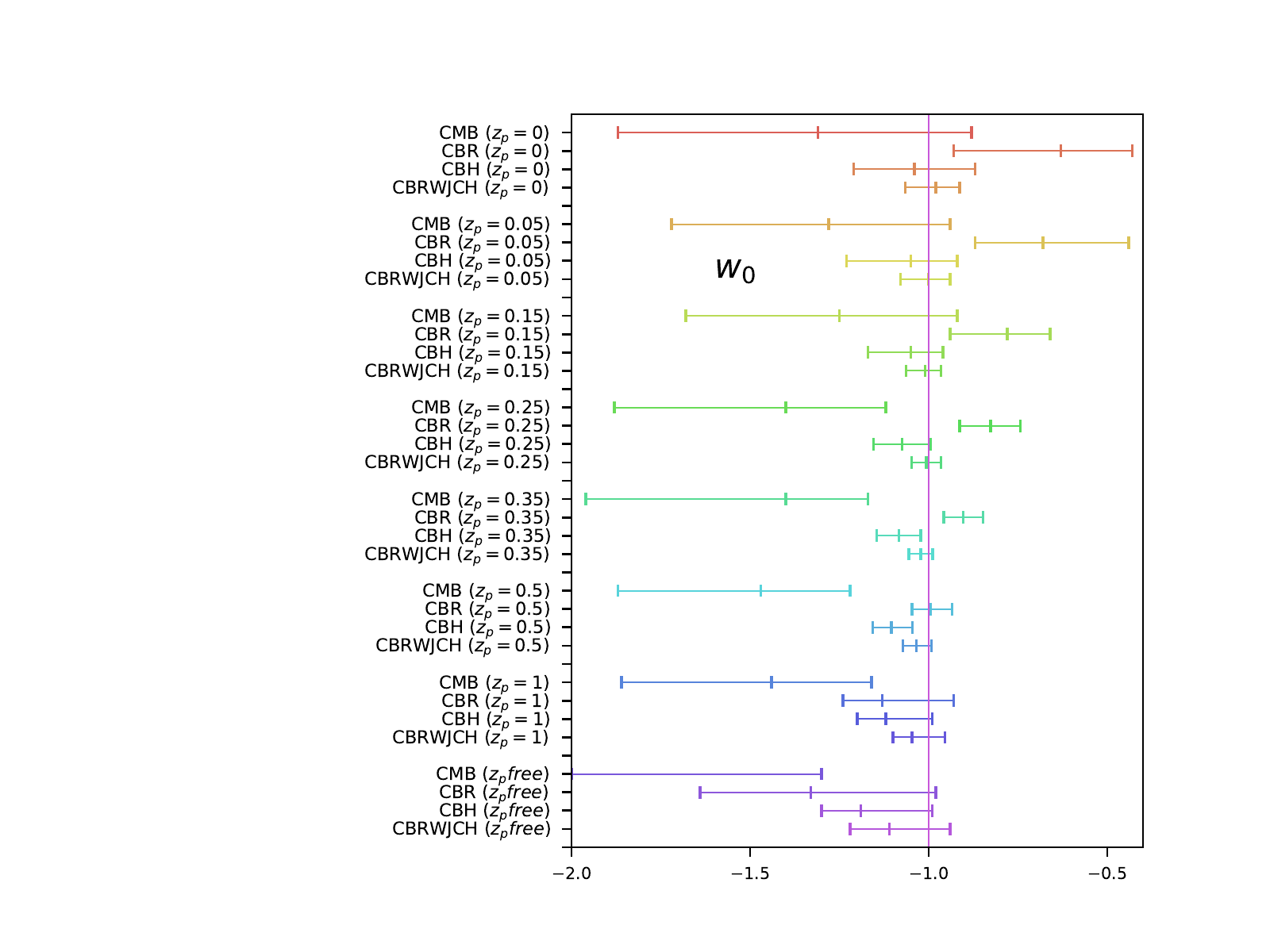}
\caption{{\it{Whisker plot for the present value of the equation-of-state parameter, $w_0$,
for all the examined cases of the generalized 
CPL parametrization (\ref{cpl-pivot}). Here, CBR = CMB+BAO+RSD, CBH = CMB+BAO+HST, and CBRWJCH = CMB+BAO+RSD+WL+JLA+CC+HST. }}}
\label{fig:whisker_w}
\end{figure*}

\section{Statistical comparison of all parametrizations}
\label{comparison}

In this section we proceed to a comparison of the various generalized CPL parametrizations, with and without the pivoting redshift. 
Hence, for completeness we perform a similar observational confrontation with the previous section for the standard CPL parametrization, namely  
without pivoting redshift (i.e. $z_p =0$), and we  summarize the results
in Table \ref{tab:cpl-no-pivot}.

We can now perform the model comparison following the summarizing Tables given
above, namely  
Table \ref{tab:varying-zp} (varying $z_p$), Table \ref{tab:zp=0.05} ($z_p =0.05$), 
Table \ref{tab:zp=0.15} ($z_p =0.15$), Table \ref{tab:zp=0.25} ($z_p =0.25$), 
Table \ref{tab:zp=0.35} ($z_p =0.35$), Table \ref{tab:zp=0.5} ($z_p =0.5$), 
  Table \ref{tab:zp=1} ($z_p =1$), and Table \ref{tab:cpl-no-pivot} (no pivoting redshift, i.e. $z_p =0$).

First of all, from all the analyses we find that the CMB data alone do not provide
stringent constraints on the free parameters, however, we observe that the addition
of any external datasets leads to a refinement of the constraints by reducing their 
error bars in a significant way. Furthermore, we find that the CMB+BAO+RSD combination
returns slightly different constraints 
compared to the remaining two datasets, nevertheless for this combination we find an
interesting pattern in the $w_0$ parameter, where we observe that with increasing $z_p$, $w_0$ eventually approaches towards the cosmological constant value and finally for large $z_p$ ($z_p = 1$) 
it crosses the $-1$ boundary.  
Concerning the remaining two datasets, namely
CMB+BAO+HST and CMB+BAO+RSD+WL+JLA+CC+HST, we find that the cosmological constraints are
similar, with the best constraints   definitely achieved for the final 
combination. 
Thus,  in this section we focus on the observational datasets CMB+BAO+RSD and 
CMB+BAO+RSD+WL+JLA+CC+HST, in order to  provide a statistical comparison
between the cosmological models for free and fixed pivoting redshift $z_p$.   

In order to proceed towards the statistical comparisons of the models, 
in Table \ref{tab:comparison} we depict the constraints on the basic model parameters
for different values of $z_p$ (free and fixed). Furthermore, in 
Fig.~\ref{fig:comparison1} we present the one-dimensional marginalized posterior distributions
for the free parameters. In particular, the upper part of Fig. \ref{fig:comparison1} 
corresponds to the CMB+BAO+RSD dataset, while the lower part to the full combination 
CMB+BAO+RSD+WL+JLA+CC+HST. From both parts of Fig. \ref{fig:comparison1} we can clearly notice that 
all parameters present  the same behavior independently of the pivoting redshift $z_p$, apart from  $w_0$ and $w_a$. 
Hence, in order to examine in more detail the effect of $z_p$ on  $w_0$,$w_a$, 
in Fig.~\ref{fig:comparison-w0-wa} we depict the contour plots in the  $w_0-w_a$  plane
for various $z_p$, for the combinations CMB+BAO+RSD (left panel of Fig.~\ref{fig:comparison-w0-wa}) and CMB+BAO+RSD+WL+JLA+CC+HST (right panel of Fig.~\ref{fig:comparison-w0-wa}). 

As we observe in Fig.~\ref{fig:comparison-w0-wa}  for both datasets,
by changing the values of $z_p$ the correlations between $w_0$ and $w_a$ 
change significantly. In particular, starting from a negative correlation 
present for $z_p=0$ (the original CPL parametrization), increasing the $z_p$ values leads
to a rotation of the direction of the degeneracy between these two parameters.
Therefore, we find a positive correlation for $z_p=1$, as well as in the case where
$z_p$ is left free.
Finally, we can identify the value of the pivoting redshift for which $w_0$ and $w_a$ 
are no more correlated, and this is approximately $z_p=0.35$ for both   datasets
(the contours corresponding to   $z_p =0.35$ (yellow) are vertical, showing no   degeneracy).
The changing of the correlations from negative to positive is one of the main results of this work. 

Lastly, in order to provide the obtained results in a more transparent way,
in Fig.~\ref{fig:whisker_w} we provide the whisker plot for the equation-of-state
parameter at present, namely $w_0$, for all the examined cases of the generalized 
CPL parametrization. As we observe, although the constraints for all data combinations 
behave in a stable way, the increase in $z_p$ pushes $w_0$ towards the phantom regime.

\section{Concluding remarks}
\label{sec-conclu}

Dynamical dark energy parametrizations are an effective approach to understand 
the evolution of the universe, without 
needing to know the microphysical origin of the dark-energy and whether it
corresponds to new fields or to gravitational modification. Hence, a large number of
such  dark-energy equation-of-state parametrizations  have 
been introduced in the literature, with the 
Chevallier-Polarski-Linder (CPL)   being one of the most studied.

Nevertheless, in most of the above  parametrizations one considers the ``pivoting redshift'' 
$z_p$ to correspond to zero, namely the  point at which the dark-energy equation-of-state 
is most tightly constrained to correspond to the current universe.
However, in the case of two-parameter models,
due to possible
rotational correlations between the two parameters, 
in principle one could avoid setting the pivoting redshift to zero
straightaway,   handling it as a free parameter.  

In the present work we investigated the observational constraints on such
a generalized CPL parametrization, namely  incorporating the 
pivoting redshift as an extra parameter, assuming it to be either fixed or free.
For this shake we used various data combinations  
from cosmic microwave background (CMB), 
baryon acoustic oscillations (BAO),
redshift space distortion (RSD), weak lensing (WL), joint light curve analysis (JLA), 
cosmic chronometers (CC), and  we additionally  included
a Gaussian prior on the Hubble constant value. We considered two different cases,
namely one in which $z_p$ is handled as a free parameter,
and one in which it is fixed to a specific value. For the later case we considered various values 
of $z_p \in [0, 1]$, in order to examine how the fixed $z_p$ value affects the results.

For the case of free $z_p$, we found that for all data combinations it
always remains unconstrained, and there is a degeneracy with the current 
value of the dark energy equation of state $w_0$ (see  Fig. \ref{w0zp}).
On the other hand, in the case where  $z_p$ is fixed
we did not find any degeneracy in the parameter space, as expected. 
In particular, the mean values of $w_0$  lie always in the phantom regime, and for 
higher values of $z_p$ ($0.25, 0.35, 0.5, 1$),  $w_0 < -1$ at more than 1$\sigma$ while
for  lower values of $z_p$ ($z_p =0, 0.05, 0.15$) the quintessence regime
is also allowed at 1$\sigma$.

The inclusion of any external data set to sole CMB data, such as BAO+RSD, BAO+HST, and
BAO+RSD+WL+JLA+CC+HST, significantly improves the CMB constraints by reducing 
the error bars on the various quantities. For instance, 
irrespectively  of the different fixed $z_p$ values, the CMB data always
return  high values for the present Hubble constant $H_0$ with large 
error bars, which both decrease for the combined data cases.

Concerning the constraints on $w_0$, for the CMB+BAO+RSD dataset 
we saw that they depend  on the values of $z_p$. 
In particular, for low $z_p$ ($z_p =0.05, 0.15, 0.25, 0.35, 0.5$)
the mean values of $w_0$ are always quintessential, while for $z_p =1$
the mean value of $w_0$ lies in the phantom regime. Nevertheless, 
a common characteristic is that for all $z_p$ values $w_0 > -1$ is allowed
within 68\% CL (note that for $z_p =0.05, 0.15, 0.25, 0.35$ within 68\% CL
$w_0$ is strictly greater than $-1$). Furthermore, for the  CMB+BAO+HST
dataset we saw that the obtained $w_0$-values
for different $z_p$ are in better agreement with the cosmological constant.
Additionally, for the last full combination of CMB+BAO+RSD+WL+JLA+CC+HST,
we also found that $w_0$ is consistent with the cosmological constant,
independently of the $z_p$ values.
As expected, compared to all the analyses performed in this work,
the cosmological constraints obtained for the full data combination, namely
CMB+BAO+RSD+WL+JLA+CC+HST, are much more stringent, as it was summarized in the 
whisker plot of  Fig.~\ref{fig:whisker_w}.

Finally, in the above analysis we were able to reveal a correlation  
between the parameters $w_0$ and $w_a$ for different $z_p$ (see Fig. \ref{fig:comparison-w0-wa}).
In particular, we found that with increasing $z_p$ the correlations between $w_0$ and $w_a$
change from negative to positive (the direction of degeneracy is rotating from negative to 
positive), and for the case $z_p =0.35$, $w_0$ and $w_a$ are uncorrelated.
This is one of the main results of the present work, and indeed it justifies why a non-zero pivoting redshift should be taken into account.

\begin{acknowledgments}
 WY has been supported by the National Natural Science Foundation of China under Grants No. 11705079 and No. 11647153. SP and WY thank G. Pantazis for many discussions. EDV acknowledges support from the European Research Council in the form of a Consolidator Grant with number 681431. This article is based upon work from CANTATA COST (European Cooperation in Science and Technology) action CA15117, 
EU Framework Programme Horizon 2020.
\end{acknowledgments}



\begin{thebibliography}{}

\bibitem{Copeland:2006wr}
  E.~J.~Copeland, M.~Sami and S.~Tsujikawa,
   {\it{Dynamics of dark energy}},
  Int.\ J.\ Mod.\ Phys.\ D {\bf 15}, 1753 (2006).
   
\bibitem{Cai:2009zp}
  Y.~F.~Cai, E.~N.~Saridakis, M.~R.~Setare and J.~Q.~Xia,
     {\it{Quintom Cosmology: Theoretical implications and observations}},
  Phys.\ Rept.\  {\bf 493}, 1 (2010).

\bibitem{modgrav1}
 S.~Nojiri and S.~D.~Odintsov,
 {\it{Introduction to modified gravity and gravitational alternative for dark energy}},
 eConf C {\bf 0602061}, 06 (2006)
  [Int.\ J.\ Geom.\ Meth.\ Mod.\ Phys.\  {\bf 4}, 115 (2007)].
  
  
  \bibitem{DeFelice:2010aj} 
  A. De Felice and S.~Tsujikawa,
  {\it $f(R)$ theories,}
  Living Rev.\ Rel.\  {\bf 13}, 3 (2010).
 
 
 \bibitem{Capozziello:2011et}
S.~Capozziello and M.~De Laurentis,
  {\it{Extended Theories of Gravity}},
Phys.\ Rept.\ {\bf 509}, 167 (2011).

\bibitem{Cai:2015emx} 
  Y.~F.~Cai, S.~Capozziello, M.~De Laurentis and E.~N.~Saridakis,
  {\it{f(T) teleparallel gravity and cosmology}},
  Rept.\ Prog.\ Phys.\  {\bf 79}, no. 10, 106901 (2016).
 
 \bibitem{Nojiri:2017ncd} 
  S.~Nojiri, S.~D.~Odintsov and V.~K.~Oikonomou,
  {\it Modified Gravity Theories on a Nutshell: Inflation, Bounce and Late-time Evolution,}
  Phys.\ Rept.\  {\bf 692}, 1 (2017).
  
\bibitem{Gong:2005de} 
  Y.~g.~Gong and Y.~Z.~Zhang,
    {\it{Probing the curvature and dark energy}},
  Phys.\ Rev.\ D {\bf 72}, 043518 (2005).
  
  \bibitem{Yang:2018qmz} 
  W.~Yang, S.~Pan, E.~Di Valentino, E.~N.~Saridakis and S.~Chakraborty,
  {\it Observational constraints on one-parameter dynamical dark-energy parametrizations and the $H_0$ tension,}
  arXiv:1810.05141 [astro-ph.CO].
  

\bibitem{Chevallier:2000qy} 
  M.~Chevallier and D.~Polarski,
    {\it{Accelerating universes with scaling dark matter}},
  Int.\ J.\ Mod.\ Phys.\ D {\bf 10}, 213 (2001).
  
  \bibitem{Linder:2002et} 
  E.~V.~Linder,
    {\it{Exploring the expansion history of the universe}},
  Phys.\ Rev.\ Lett.\  {\bf 90}, 091301 (2003).

  
 \bibitem{Cooray:1999da} 
  A.~R.~Cooray and D.~Huterer,
    {\it{Gravitational lensing as a probe of quintessence}},
  Astrophys.\ J.\  {\bf 513}, L95 (1999).

   
 \bibitem{Astier:2000as} 
  P.~Astier,
    {\it{Can luminosity distance measurements probe the equation of state of dark 
energy}},
  Phys.\ Lett.\ B {\bf 500}, 8 (2001).
    
  
\bibitem{Weller:2001gf} 
  J.~Weller and A.~Albrecht,
    {\it{Future supernovae observations as a probe of dark energy}},
  Phys.\ Rev.\ D {\bf 65}, 103512 (2002).


\bibitem{Efstathiou:1999tm} 
  G.~Efstathiou,
    {\it{Constraining the equation of state of the universe from distant type Ia 
supernovae 
and cosmic 
microwave background anisotropies}},
  Mon.\ Not.\ Roy.\ Astron.\ Soc.\  {\bf 310}, 842 (1999).
 


\bibitem {Jassal:2005qc}
H.~K.~Jassal, J.~S.~Bagla and T.~Padmanabhan,
  {\it{Observational constraints on low redshift evolution of dark energy: How consistent are different observations?}},
Phys.\ Rev.\ D \textbf{72}, 103503 (2005).
 


\bibitem {Barboza:2008rh}E.~M.~Barboza, Jr. and J.~S.~Alcaniz,
  {\it{A parametric model for dark energy}},
Phys.\ Lett.\ B \textbf{666}, 415 (2008). 



  
\bibitem{Ma:2011nc} 
  J.~Z.~Ma and X.~Zhang,
    {\it{Probing the dynamics of dark energy with novel parametrizations}},
  Phys.\ Lett.\ B {\bf 699}, 233 (2011).
 
   

 
\bibitem{Nesseris:2004wj} 
  S.~Nesseris and L.~Perivolaropoulos,
    {\it{A Comparison of cosmological models using recent supernova data}},
  Phys.\ Rev.\ D {\bf 70}, 043531 (2004).
            
            

\bibitem{Linder:2005ne} 
  E.~V.~Linder and D.~Huterer,
    {\it{How many dark energy parameters?}},
  Phys.\ Rev.\ D {\bf 72}, 043509 (2005).

            
 
  \bibitem{Feng:2004ff} 
  B.~Feng, M.~Li, Y.~S.~Piao and X.~Zhang,
    {\it{Oscillating quintom and the recurrent universe}},
  Phys.\ Lett.\ B {\bf 634}, 101 (2006).
 

\bibitem{Zhao:2006qg}
  G.~B.~Zhao, J.~Q.~Xia, H.~Li, C.~Tao, J.~M.~Virey, Z.~H.~Zhu and X.~Zhang,
  {\it{Probing for dynamics of dark energy and curvature of universe with latest cosmological observations}},
  Phys.\ Lett.\ B {\bf 648}, 8 (2007).
    
 
\bibitem{Nojiri:2006ww} 
  S.~Nojiri and S.~D.~Odintsov,
  {\it{The Oscillating dark energy: Future singularity and coincidence problem}},
  Phys.\ Lett.\ B {\bf 637}, 139 (2006). 
  
  
 \bibitem{Saridakis:2008fy} 
  E.~N.~Saridakis,
{\it{Theoretical Limits on the Equation-of-State Parameter of Phantom Cosmology}},
  Phys.\ Lett.\ B {\bf 676}, 7 (2009).
 
   
\bibitem{Dutta:2009yb} 
  S.~Dutta, E.~N.~Saridakis and R.~J.~Scherrer,
 {\it{Dark energy from a quintessence (phantom) field rolling near potential minimum 
(maximum)}},
  Phys.\ Rev.\ D {\bf 79}, 103005 (2009).
  
  
\bibitem{Lazkoz:2010gz} 
  R.~Lazkoz, V.~Salzano and I.~Sendra,
    {\it{Oscillations in the dark energy EoS: new MCMC lessons}},
  Phys.\ Lett.\ B {\bf 694}, 198 (2011).

 
  
 
 \bibitem{Feng:2011zzo} 
  L.~Feng and T.~Lu,
    {\it{A new equation of state for dark energy model}},
  JCAP {\bf 1111}, 034 (2011).

 
\bibitem{Saridakis:2009pj} 
  E.~N.~Saridakis,
{\it{Phantom evolution in power-law potentials}},
  Nucl.\ Phys.\ B {\bf 819}, 116 (2009).

\bibitem{DeFelice:2012vd} 
  A.~De Felice, S.~Nesseris and S.~Tsujikawa,
    {\it{Observational constraints on dark energy with a fast varying equation of state}},
  JCAP {\bf 1205}, 029 (2012).
  
  
\bibitem{Saridakis:2009ej} 
  E.~N.~Saridakis,
{\it{Quintom evolution in power-law potentials}},
  Nucl.\ Phys.\ B {\bf 830}, 374 (2010).
  
  
  
  \bibitem{Feng:2012gf} 
  C.~J.~Feng, X.~Y.~Shen, P.~Li and X.~Z.~Li,
    {\it{A New Class of Parametrization for Dark Energy without Divergence}},
  JCAP {\bf 1209}, 023 (2012).
  
 \bibitem{Basilakos:2013vya} 
  S.~Basilakos and J.~Sol\'{a},
  {\it{ Effective equation of state for running vacuum: ‘mirage’ quintessence and phantom 
dark energy}},
  Mon.\ Not.\ Roy.\ Astron.\ Soc.\  {\bf 437}, no. 4, 3331 (2014).
 
 \bibitem{Pantazis:2016nky} 
  G.~Pantazis, S.~Nesseris and L.~Perivolaropoulos,
    {\it{Comparison of thawing and freezing dark energy parametrizations}},
  Phys.\ Rev.\ D {\bf 93}, no. 10, 103503 (2016).

 \bibitem{DiValentino:2016hlg} 
  E.~Di Valentino, A.~Melchiorri and J.~Silk,
  {\it{Reconciling Planck with the local value of $H_0$ in extended parameter space}},
  Phys.\ Lett.\ B {\bf 761}, 242 (2016).
  
  
\bibitem{Chavez:2016epc} 
  R.~Ch\'{a}vez, M.~Plionis, S.~Basilakos, R.~Terlevich, E.~Terlevich, J.~Melnick, F.~Bresolin and 
A.~L.~Gonz\'{a}lez-Mor\'{a}n,
  {\it {Constraining the dark energy equation of state with H $\small{II}$ galaxies}},
  Mon.\ Not.\ Roy.\ Astron.\ Soc.\  {\bf 462}, no. 3, 2431 (2016).
  
  
  
  
 \bibitem{Zhao:2017cud}
  G.~B.~Zhao {\it et al.},
  {\it{Dynamical dark energy in light of the latest observations}},
  Nat.\ Astron.\  {\bf 1}, 627 (2017).
  
  
\bibitem{Yang:2017amu} 
  W.~Yang, R.~C.~Nunes, S.~Pan and D.~F.~Mota,
  {\it{Effects of neutrino mass hierarchies on dynamical dark energy models}},
  Phys.\ Rev.\ D {\bf 95}, 103522 (2017). 
  
  
\bibitem{DiValentino:2017zyq} 
  E.~Di Valentino, A.~Melchiorri, E.~V.~Linder and J.~Silk,
  {\it{Constraining Dark Energy Dynamics in Extended Parameter Space}},
  Phys.\ Rev.\ D {\bf 96},  023523 (2017).
 
  

\bibitem{DiValentino:2017gzb} 
  E.~Di Valentino,
  {\it{Crack in the cosmological paradigm}},
  Nat.\ Astron.\  {\bf 1}, 569 (2017).
 

  
\bibitem{Yang:2017alx} 
  W.~Yang, S.~Pan and A.~Paliathanasis,
    {\it{Latest astronomical constraints on some nonlinear parametric dark energy 
models}}, Mon.\ Not.\ Roy.\ Astron.\ Soc.\  {\bf 475}, 2605 (2018).
 
 
\bibitem{Marcondes:2017vjw} 
  R.~J.~F.~Marcondes and S.~Pan,
  {\it{Cosmic chronometers constraints on some fast-varying dark energy equations of 
state}}, arXiv:1711.06157 [astro-ph.CO]. 

  
 
  \bibitem{Pan:2017zoh} 
  S.~Pan, E.~N.~Saridakis and W.~Yang,
  {\it Observational Constraints on Oscillating Dark-Energy Parametrizations,}
  Phys.\ Rev.\ D {\bf 98}, no. 6, 063510 (2018).
  
  

\bibitem{Linder:2006xb} 
  E.~V.~Linder,
  {\it Biased Cosmology: Pivots, Parameters, and Figures of Merit,}
  Astropart.\ Phys.\  {\bf 26}, 102 (2006).
  
  
    
\bibitem{Sahni:2006pa} 
  V.~Sahni and A.~Starobinsky,
  {\it Reconstructing Dark Energy,}
  Int.\ J.\ Mod.\ Phys.\ D {\bf 15}, 2105 (2006).
  

  
  \bibitem{Scovacricchi:2014haa} 
  D.~Scovacricchi, S.~A.~Bonometto, M.~Mezzetti and G.~La Vacca,
  {\it Constraints on Dark Energy state equation with varying pivoting redshift,}
  New Astron.\  {\bf 26}, 106 (2014). 


 
\bibitem{Mukhanov}V. F. Mukhanov, H. A. Feldman and R. H. Brandenberger,
{\it Theory of cosmological perturbations},
Phys. Rept. \textbf{215}, 203 (1992).

\bibitem {Ma:1995ey}C.~P.~Ma and E.~Bertschinger,
{\it Cosmological perturbation theory in the synchronous and conformal Newtonian gauges,}
Astrophys.\ J.\ \textbf{455}, 7 (1995).



\bibitem{Malik:2008im} 
  K.~A.~Malik and D.~Wands,
  {\it Cosmological perturbations,}
  Phys.\ Rept.\  {\bf 475}, 1 (2009).
  


\bibitem{Adam:2015rua} 
  R.~Adam {\it et al.} [Planck Collaboration],
    {\it{Planck 2015 results. I. Overview of products and scientific results}},
  Astron.\ Astrophys.\  {\bf 594}, A1 (2016).
  
  \bibitem{Aghanim:2015xee} 
  N.~Aghanim {\it et al.} [Planck Collaboration],
    {\it{Planck 2015 results. XI. CMB power spectra, likelihoods, and robustness of 
parameters}},
  Astron.\ Astrophys.\  {\bf 594}, A11 (2016).
  
  \bibitem{Akrami:2018vks} 
  Y.~Akrami {\it et al.} [Planck Collaboration],
  {\it Planck 2018 results. I. Overview and the cosmological legacy of Planck,}
  arXiv:1807.06205 [astro-ph.CO].
 
  
  
\bibitem{planckparams2018} 
  N.~Aghanim {\it et al.} [Planck Collaboration],
  {\it Planck 2018 results. VI. Cosmological parameters,}
  arXiv:1807.06209 [astro-ph.CO].
  
  
  \bibitem{planckparams2015}
  P.~A.~R.~Ade {\it et al.}  [Planck Collaboration],
  {\it Planck 2015 results. XIII. Cosmological parameters,}
  arXiv:1502.01589 [astro-ph.CO].
  

\bibitem{Beutler:2011hx} 
  F.~Beutler {\it et al.},
    {\it{The 6dF Galaxy Survey: Baryon Acoustic Oscillations and the Local Hubble 
Constant}},
  Mon.\ Not.\ Roy.\ Astron.\ Soc.\  {\bf 416}, 3017 (2011).
       
\bibitem{Ross:2014qpa} 
  A.~J.~Ross, L.~Samushia, C.~Howlett, W.~J.~Percival, A.~Burden and M.~Manera, {\it{The clustering of the SDSS DR7 main Galaxy sample I. A 4 per cent distance measure at $z = 0.15$}},
  Mon.\ Not.\ Roy.\ Astron.\ Soc.\  {\bf 449}, no. 1, 835 (2015).
          
\bibitem{Gil-Marin:2015nqa} 
  H.~Gil-Mar\'{i}n {\it et al.},
    {\it{The clustering of galaxies in the SDSS-III Baryon Oscillation Spectroscopic Survey: BAO measurement from the LOS-dependent power spectrum of DR12 BOSS galaxies}},
  Mon.\ Not.\ Roy.\ Astron.\ Soc.\  {\bf 460}, no. 4, 4210 (2016).
            
\bibitem{Gil-Marin:2016wya}
  H.~Gil-Mar\'{i}n {\it et al.},
 {\it{The clustering of galaxies in the SDSS-III Baryon Oscillation Spectroscopic Survey: RSD measurement from the power spectrum and bispectrum of the DR12 BOSS galaxies}},
  Mon.\ Not.\ Roy.\ Astron.\ Soc.\  {\bf 465}, no.2,  1757 (2017). 
  
  
  \bibitem {Heymans:2013fya}
C.~Heymans \textit{et al.},
  {\it{CFHTLenS tomographic weak lensing cosmological parameter constraints: Mitigating the impact of 
intrinsic galaxy alignments}},
Mon.\ Not.\ Roy.\ Astron.\ Soc.\ \textbf{432}, 2433 (2013).
 
\bibitem{Erben:2012zw} 
  T.~Erben {\it et al.},
    {\it{CFHTLenS: The Canada-France-Hawaii Telescope Lensing Survey - Imaging Data and Catalogue Products}},
  Mon.\ Not.\ Roy.\ Astron.\ Soc.\  {\bf 433}, 2545 (2013).
  
   
\bibitem {Asgari:2016xuw}
M.~Asgari, C.~Heymans, C.~Blake, J.~Harnois-Deraps,
P.~Schneider and L.~Van Waerbeke,
  {\it{Revisiting CFHTLenS cosmic shear: Optimal E/B mode decomposition using COSEBIs and compressed COSEBIs}},
Mon.\ Not.\ Roy.\ Astron.\ Soc.\ \textbf{464}, 1676 (2017).


\bibitem{Betoule:2014frx} 
  M.~Betoule {\it et al.} [SDSS Collaboration],
    {\it{Improved cosmological constraints from a joint analysis of the SDSS-II and SNLS 
supernova 
samples}},
  Astron.\ Astrophys.\  {\bf 568}, A22 (2014).
       
\bibitem {Moresco:2016mzx}
M.~Moresco \textit{et al.},
  {\it{A 6\% measurement of the Hubble parameter at $z\sim0.45$: direct evidence of the 
epoch 
of cosmic 
re-acceleration}},
JCAP \textbf{1605}, 014 (2016).
 
\bibitem{Riess:2016jrr} 
  A.~G.~Riess {\it et al.},
    {\it{A 2.4\% Determination of the Local Value of the Hubble Constant}},
  Astrophys.\ J.\  {\bf 826}, no. 1, 56 (2016).

\bibitem{Lewis:2002ah}
  A.~Lewis and S.~Bridle,
  \textit{``Cosmological parameters from CMB and other data: A Monte Carlo approach,''}
  Phys.\ Rev.\ D {\bf 66}, 103511 (2002). 
  
  \bibitem{Gelman-Rubin}
  A.~Gelman and D.~Rubin, 
  {\it Inference from iterative simulation using multiple sequences,} 
  Statistical Science \textbf{7}, 457 (1992).

 
\bibitem{Lewis:2013hha} 
  A.~Lewis,
  {\it Efficient sampling of fast and slow cosmological parameters,}
  Phys.\ Rev.\ D {\bf 87}, no. 10, 103529 (2013). 
  

\bibitem{Ade:2015xua}
  P.~A.~R.~Ade {\it et al.} [Planck Collaboration],
  {\it Planck 2015 results. XIII. Cosmological parameters,}
  Astron.\ Astrophys.\  {\bf 594}, A13 (2016). 



\end{thebibliography}
\end{document}